\documentclass[journal]{IEEEtran}
\usepackage{cite}
\usepackage[cmex10]{amsmath}
\usepackage{amsfonts}
\usepackage{amssymb}
\usepackage{array}
\usepackage{caption}
\usepackage{booktabs}
\usepackage{tabularx}
\usepackage{ragged2e}
\usepackage{graphicx}
\graphicspath{{figures/}}
\usepackage{stfloats}
\usepackage{placeins}
\usepackage{rotating}
\usepackage{url}
\usepackage{microtype}
\usepackage{xcolor}
\usepackage{hyperref}
\newcolumntype{L}[1]{>{\RaggedRight\arraybackslash}p{#1}}
\newcolumntype{C}[1]{>{\Centering\arraybackslash}p{#1}}
\newcolumntype{Y}{>{\RaggedRight\arraybackslash}X}
\newcolumntype{Z}[1]{>{\hsize=#1\hsize\linewidth=\hsize\RaggedRight\arraybackslash}X}
\hypersetup{
    hidelinks,
    pdftitle={Towards channel foundation models (CFMs): Motivations, methodologies and opportunities},
    pdfauthor={Jun Jiang, Yuan Gao, Xinyi Wu, and Shugong Xu},
    pdfkeywords={channel foundation models, wireless channel, channel state information, representation learning, pretraining, transfer learning}
}
\begin{document}

\title{Towards channel foundation models (CFMs): Motivations, methodologies and opportunities}

\author{Jun Jiang, Yuan Gao, Xinyi Wu, Shugong Xu, \textit{Fellow, IEEE}

\thanks{Jun Jiang and Yuan Gao contributed equally. Corresponding author: Shugong Xu}
\thanks{Jun Jiang and Shugong Xu are with the School of Advanced Technology, Xi'an Jiaotong-Liverpool University, Suzhou, China (e-mail: jun.jiang25@student.xjtlu.edu.cn and shugong.xu@xjtlu.edu.cn).}
\thanks{Yuan Gao and Xinyi Wu are with the School of Communication and Information Engineering, Shanghai University, Shanghai, China (e-mail: gaoyuansie@shu.edu.cn; wu\_xinyi0312@shu.edu.cn).}
}

\maketitle

\begin{abstract}
Artificial intelligence (AI) models for wireless systems are commonly optimized for fixed tasks, channel distributions, and system configurations, which limits their reuse when downstream objectives or operating conditions change. Channel foundation models (CFMs) address this limitation by pretraining reusable models on channel-informative observations and adapting their parameters or representations across channel-related tasks and conditions. However, existing studies differ substantially in their input modalities, pretraining targets, retained components, and transfer settings, making it difficult to compare how the methods are constructed, reused, and evaluated. This paper surveys CFM research through the central question of how channel-informative pretraining is converted into reusable downstream capabilities. We adopt a channel-centered scope covering explicit channel state information, received-signal observations, and auxiliary environmental or geometric context. Existing methods are organized according to the source of their learning targets into observation-space modeling, relational representation learning, latent-space prediction, task-target learning, system-utility optimization, and hybrid combinations. For representative methods, we compare how reusable parameters are learned and adapted, their downstream applications, and the settings used to evaluate transfer. The resulting framework connects CFM methodologies to physical-layer processing, radio access network and medium access control tasks, integrated sensing and communication, and radio-environment modeling. A matched case study further compares transfer for channel extrapolation, beam prediction, and positioning under common backbone and source-data settings. Finally, we discuss open problems in data construction, transfer evaluation, efficient adaptation, deployment, reliability, privacy, and security.
\end{abstract}


\begin{figure}[!t]
\centering
\includegraphics[
    width=0.85\columnwidth,
    trim={96bp 580bp 167bp 111bp},
    clip
]{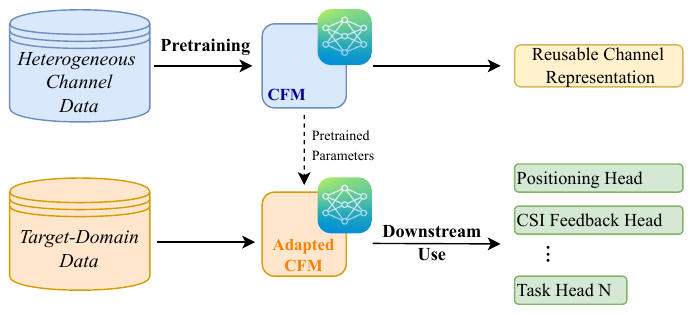}
\caption{Pretraining and downstream adaptation of a channel foundation model (CFM). Heterogeneous channel data are used to learn a reusable channel representation. The pretrained parameters initialize a CFM that is adapted using target-domain data and connected to downstream task heads.}
\label{fig:overall}\vspace{-4mm}
\end{figure}
\IEEEpeerreviewmaketitle
\section{Introduction}
Learning-based methods have been applied to channel estimation, signal detection, channel state information (CSI) feedback, beam management, and sensing~\cite{oshea2017introduction,8640815,10597358,xue2024beammanagement}. These models are commonly optimized for a fixed input-output mapping and a fixed training distribution. Variations in the propagation environment, carrier frequency, bandwidth, antenna configuration, and mobility induce a mismatch between the training condition and the deployment condition. Release~18 examined artificial intelligence and machine learning (AI/ML) for the New Radio (NR) air interface under several task- and configuration-specific settings~\cite{3gpp38843}. The reported studies therefore remain bound to a prescribed task and radio configuration. The open question is whether pretrained channel parameters or representations can be retained when the downstream task or the operating condition changes.

Reuse is difficult for a concrete physical reason rather than because each task uses a different output head. The wireless channel is the shared origin of the shift. The same multipath process determines pilot observations, payload measurements, beam responses, sensing echoes, etc. A change in site geometry, carrier frequency, or array configuration therefore perturbs multiple downstream problems at once. Acquisition of channel observations is already costly, and the construction of task labels or reference channel quantities further increases the measurement burden. A model trained for one task-condition pair is frequently replaced, after which the associated data collection and optimization procedures are repeated~\cite{10288574}.

If channel structure must be relearned for every new pair, a direct alternative is to pretrain a reusable block once and adapt it later. That alternative is the workflow associated with foundation models (FMs) in natural language processing (NLP) and computer vision (CV)~\cite{bommasani2021opportunities,devlin2019bertpretrainingdeepbidirectional,radford2021clip,he2022masked}. Broadly pretrained parameters are reused through prompting, task heads, or restricted parameter updates, instead of training a separate model for each target. Transfer performance still depends on the pretraining target, the coverage of the pretraining corpus, the retained component, and the downstream interface. Model scale alone does not determine the outcome. For wireless systems the same workflow is useful only if the retained block encodes the quantity that actually shifts across tasks and deployments, which is channel structure.

Current research on wireless communications does not isolate that block. A task-specific model constitutes a complete mapping for one condition. Multitask learning shares parameters across a predefined output set, yet joint training on that set does not establish reuse for an unseen task or an unseen channel condition~\cite{9392366}. A general-purpose language or vision model may be adapted to a channel problem after CSI or a received waveform is converted into tokens or image-compatible patches. The base representation is nevertheless learned from text or natural images rather than from electromagnetic structure. The designation \textit{wireless foundation model} (WFM) has been applied to all of the above settings, and also to studies in which the wireless link serves only as a transport medium for features or model updates. These usages are separated in Section~\ref{subsec:related}.

Channel foundation models (CFMs) keep the FM workflow while placing pretraining on channel-informative observations~\cite{bommasani2021opportunities,xu2026nativeaicfm}. An encoder, backbone, generative prior, or equivalent parameter block is first pretrained on such observations, retained as the CFM, and subsequently applied or adapted to a channel-related task or operating condition. Fig.~\ref{fig:overall} illustrates this two-stage procedure. In this survey, a study is classified as a CFM work only if the channel-pretrained block is retained and reuse is targeted beyond a predetermined task-condition pair. Transfer learning provides the adaptation mechanism. A single pretraining and fine-tuning pipeline on a fixed pair is not a sufficient criterion.

CFM pretraining is not synonymous with self-supervision. Learning targets may be formed from withheld observations, relations among views, latent features, supervised task outputs, differentiable system utilities, or combinations of these signals. The observations may comprise explicit CSI, received in-phase/quadrature (IQ) samples, spectra, pilots, and auxiliary environmental measurements that retain channel effects. Section~III organizes existing methods according to the source of the target used to update the retained parameters.

\subsection{Related Work}
\label{subsec:related}
The models discussed above are not named consistently in the literature, and the naming directly affects the scope of a survey. Prior reviews have covered generative and self-supervised learning, large language and multimodal models, and wireless FMs~\cite{celik2024dawn,10930396,boateng2025survey,zhou2024large,liang2026foundationwireless}. For channel-related problems, two lines of work can be distinguished. One line adapts a general-purpose language model to a channel task such as extrapolation~\cite{liu2024llm4cp}. The other pretrains models directly on CSI or received-signal observations~\cite{gao2026ai,gao2025enabling,buffelli2025foundationcommunication}.

The term WFM does not separate these lines. Several studies employ it for models pretrained on wireless observations, including WiFo for CSI-based channel extrapolation~\cite{liu2025wifo}, the model of Y.~Sheng \textit{et al.} for channel-sequence prediction together with integrated sensing and communication (ISAC) angle-of-arrival (AoA) and traffic prediction~\cite{sheng2025wirelessfoundationmodelmultitask}, SpectrumFM for received-signal analysis~\cite{zhou2025spectrumfm}, the radio model of J.~Fontaine \textit{et al.}~\cite{fontaine2024towards}, the spectrogram model of A.~Aboulfotouh \textit{et al.}~\cite{aboulfotouh2024building}, and IQFM for received-signal representation learning~\cite{mashaal2025iqfm}. These models constitute the closest antecedents of the CFMs reviewed in this survey.

The same designation also covers studies whose base models are not pretrained on wireless observations. General-purpose language or vision models have been incorporated into semantic image transmission, where the pretrained representation characterizes source content rather than the radio channel~\cite{zhao2024lamosc,ma2025largeaisemantic}. Other works address semantic communication and language-model inference over wireless links~\cite{jin2025joint}, as well as federated fine-tuning and over-the-air-assisted adaptation of pretrained models~\cite{wang2024federatedfinetuning,chen2026airbornefm}. In the latter setting, the wireless link primarily transports intermediate features or parameter updates. Broader surveys further extend the scope to resource allocation, semantic communication, and autonomous network management~\cite{liang2026foundationwireless}. WFM is therefore used as a domain designation that spans heterogeneous pretraining data, model roles, and reuse objectives.

This survey instead adopts CFM as the organizing term. The wireless channel provides a well-defined object of pretraining because it determines received-signal formation and enters communication and sensing functions throughout the system. The CFM scope comprises models that pretrain and reuse representations derived from explicit CSI, or from received signals and auxiliary context that contain channel effects. Visual or geometric observations may be included as complementary inputs. They do not substitute for channel-informative observations. Within this scope, we compare pretraining objectives, retained components, adaptation protocols, downstream tasks, and evaluation settings.

A continuously updated collection of the reviewed CFM papers and related resources is maintained at \href{https://github.com/GREAT-ISAC/Awesome-Channel-Foundation-Models}{\textcolor{blue}{\textit{Awesome-Channel-Foundation-Models}}}\footnote{\url{https://github.com/GREAT-ISAC/Awesome-Channel-Foundation-Models}}.
\subsection{Contributions and Scope}
The survey makes three contributions.
\begin{itemize}
    \item \textbf{Conceptual positioning and scope.} We define CFMs by channel-informative pretraining, retention of the resulting parameter block or representation, and reuse across a new task or a new operating condition. This definition distinguishes CFMs from task-specific models, from multitask models trained on a closed output set, and from general-purpose FMs adapted to wireless inputs. 
    \item \textbf{Research landscape and methodological comparison.} Existing studies are compared in terms of inputs, retained parameters or representations, pretraining objectives, adaptation protocols, downstream tasks, and evaluation settings. Pretraining methods are organized by the target used to update the reusable component, namely observation-space modeling, relational representation learning, latent-space prediction, task-target learning, and system-utility optimization. Hybrid methods combine targets from two or more groups.
    \item \textbf{Applications, evaluation, and open problems.} The surveyed methods are connected to physical-layer (PHY), radio access network (RAN), medium access control (MAC), ISAC, and radio-environment tasks. A matched case study compares transfer under a common backbone and source dataset. The cross-method comparison is used to identify limitations in data construction, transfer evaluation, deployment cost, reliability, privacy, and security. It is not used to rank papers trained under incompatible experimental protocols.
\end{itemize}

Section~II defines CFMs and downstream adaptation. Section~III reviews data, architectures, pretraining objectives, adaptation, efficiency, and cross-method trade-offs. Section~IV connects these methods to applications and evaluation, Section~V identifies open problems, and Section~VI concludes the survey.

\begin{table*}[t]
\centering
\caption{Comparison of model-construction and reuse workflows for channel-related tasks.}
\label{tab:model_comparison}
\small
\setlength{\tabcolsep}{3pt}
\renewcommand{\arraystretch}{1.15}
\begin{tabular*}{\textwidth}{@{}l@{\extracolsep{\fill}}llll@{}}
\toprule
\textbf{Aspect} & \textbf{Task-specific} & \textbf{Multitask} & \textbf{Adapted general FM} & \textbf{CFM} \\ \midrule
Base-model goal & Target task & Fixed task set & General capabilities & Channel capabilities \\
Channel learning & Task training & Joint training & Downstream transfer & Base pretraining \\
Reused part & Task model & Shared module & General pretrained base & Channel pretrained base \\
Downstream use & Task inference & Joint task inference & Inference or adaptation & Inference or adaptation \\
Intended scope & Task--condition pair & Predefined tasks & Transfer to channel tasks & Channel tasks/conditions \\ \bottomrule
\end{tabular*}\vspace{-4mm}
\end{table*}

\section{Motivation, Scope, and Adaptation of CFMs}

\subsection{Why CFMs for Channel-Related Tasks}
Section~I identified task-condition binding as the operational limitation of current wireless learning models. The present section specifies the channel-centered scope that follows from that limitation and the criteria used to decide whether a pretrained model belongs to the CFM class.

The wireless channel determines the received signal and enters both communication and sensing functions. Standardized and geometry-based models describe how this channel varies across environments and radio configurations~\cite{3gpp38901,jaeckel2014quadriga}. Received observations are not equivalent to the channel itself. They also contain the transmitted waveform, interference, radio-frequency hardware, and the signal-processing chain used at acquisition. Measurements collected under one combination of these factors can therefore depart from the observations encountered in deployment.

That departure is the reason for constructing a channel-pretrained base. Collecting measurements over many environments and radio configurations is costly. Task labels and reference channel quantities add further measurement or simulation effort. Pretraining on diverse channel-informative observations is intended to extract information that remains usable when the downstream task or the operating condition changes. The same argument does not imply that every wireless function should be realized as a CFM. Any claimed benefit is testable only through adaptation and evaluation under conditions that differ from pretraining.
\subsection{How CFMs Differ from Related Models}
The motivation above still leaves CFMs adjacent to several other model families. Section~I-A treated WFM as a domain designation rather than a single model class. CFMs occupy a narrower position inside that designation. They learn channel-related structure and dependencies during base-model pretraining and retain the resulting component for later channel tasks or operating conditions. The deciding factors are the object of pretraining and the intended reuse. Model name, architecture, scale, and protocol layer do not determine membership. A channel-focused model may still be called a WFM. Many models published under that name address source content, network management, or distributed learning rather than channel reuse.

Table~\ref{tab:model_comparison} compares four construction routes used for channel-related tasks. The routes are task-specific modeling, multitask modeling, adaptation of a general-purpose FM, and construction of a CFM. These workflows can overlap in implementation. The table compares roles and does not partition the literature into disjoint classes.

Foundation-model use is identified by the scope of reuse, not by the absence of a concrete training objective. CFM pretraining may optimize reconstruction, extrapolation, or supervised task losses. Masked CSI reconstruction implements extrapolation when the mask covers a designated time, frequency, or antenna region. Random masking more generally trains completion~\cite{liu2025wifo,jiang2026csimaemaskedautoencoderbasedchannel}. The pretraining objective and a later downstream objective may therefore be identical. The distinguishing requirement is that the retained component function as a reusable base beyond one fixed task-condition pair. Task training alone does not create that base. A pretraining and fine-tuning pipeline confined to one pair does not create it either.

The two FM routes differ in where channel capability is acquired. An adapted general-purpose FM starts from a representation trained on text or natural images and transfers it into a channel application. LVM4CSI is an example. It maps CSI to a vision-compatible input~\cite{guo2025lvm4csi}. A CFM instead places base-model pretraining on channel structure and dependencies so that the retained block can be reused across channel tasks or conditions. The routes are not isolated from each other. A general-purpose FM may become a CFM after channel-domain continued pretraining and after the resulting block is reused across channel tasks or conditions. A distinctive architecture, a distinctive loss, or the ability to perform direct inference does not by itself mark the difference. Both routes permit direct inference when the pretrained interface already matches the target. Otherwise both require adaptation. Generalization and computational cost remain properties of the chosen model, data, adaptation protocol, and deployment setting.

\subsection{Definition and Key Features of CFMs}
The distinctions in Table~\ref{tab:model_comparison} are collected in an operational definition. A CFM is a reusable base model pretrained on channel-related observations. Its pretrained parameters or learned representations are retained and then applied or adapted across channel-related tasks or operating conditions. They are not used only as the final model of one fixed task-condition pair. The reused part may be an encoder, a backbone, a generative model, or another pretrained module. No particular architecture, parameter count, or pretraining objective is required~\cite{bommasani2021opportunities,vaswani2017attention}. Architecture and scale do not waive the three conditions that follow.

The first condition concerns the input. A CFM must receive channel-informative observations, either explicit CSI or received signals that still contain channel effects. Visual or geometric measurements may supply auxiliary context. They do not define a CFM in the absence of channel-related input. For received-signal models, the learned representation must retain and reuse channel information. Models that suppress channel effects in order to recognize waveforms or devices pursue a different objective and fall outside the present scope.

The second condition is retention of a channel-pretrained component after pretraining. The third condition is reuse in at least one of two forms. \textit{Cross-task reuse} introduces a task that was absent from pretraining. \textit{Cross-condition reuse} keeps the task and changes the channel environment or the radio configuration~\cite{pan2010transfer}. Joint training on a predetermined task set does not constitute cross-task reuse. This survey includes a study only if it retains a channel-pretrained base and examines reuse for a new task, a materially different channel condition, or both. A one-off pretraining and fine-tuning run on a predetermined task-condition pair is not sufficient. Transfer learning is the mechanism through which reuse is realized. It is not a sufficient CFM criterion. The demonstrated breadth is exactly the set of tasks and conditions that were evaluated. Section~\ref{sec:adaptation} specifies how the retained parameters are used in those evaluations.

\subsection{How a Pretrained CFM Is Used Downstream}
\label{sec:adaptation}
The inclusion rule above is incomplete without an account of how the retained block is probed. The adaptation protocol is not an implementation detail. It determines what a downstream result can be said to prove about the pretrained component.

After pretraining, a CFM is used in five ways that differ by which parameters are updated and by whether target-task training is required. Full fine-tuning updates the pretrained model and its task head. Linear probing and head-only adaptation keep the backbone fixed and train a linear or nonlinear output module, respectively. Parameter-efficient fine-tuning (PEFT) updates only a small internal or attached parameter set, such as a low-rank adaptation (LoRA) module~\cite{hu2022lora}. Direct inference updates no parameter and is available only when the pretrained output already matches the target. Table~\ref{tab:adaptation} summarizes these choices.

Each protocol tests a different claim. Linear probing asks whether the required information is linearly accessible in a frozen representation. A nonlinear head can add a further mapping on top of that representation. PEFT modifies a restricted part of the feature transformation. Full fine-tuning mainly tests whether pretraining supplies a useful initialization under a larger update budget. A low pretraining loss does not, on its own, demonstrate downstream accuracy or data efficiency.

An informative evaluation reports the downstream metric, the target-data budget, the number of trainable parameters, the distribution shift, and the computational cost, together with a matched from-scratch model~\cite{10288574}. In-distribution adaptation measures reuse under familiar conditions. Held-out-condition adaptation measures performance after exposure to a new condition. Direct inference measures reuse without target-task training. These settings should be reported separately. Reducing the number of updated parameters also does not guarantee lower inference latency or memory consumption, because the frozen backbone remains in the forward path.
\begin{table*}[t]
\caption{Downstream adaptation and direct-use strategies for a pretrained CFM\@.}
\label{tab:adaptation}
\centering
\footnotesize
\setlength{\tabcolsep}{3.5pt}
\renewcommand{\arraystretch}{1.12}
\begin{tabularx}{\textwidth}{@{}L{2.9cm}L{2.3cm}L{3.8cm}C{2.4cm}Y@{}}
\toprule
\textbf{Strategy} & \textbf{Update to $\boldsymbol{\theta}_0$} & \textbf{Added or updated part} & \textbf{Target training} & \textbf{Interpretation} \\ \midrule
Full fine-tuning & All & Task head or decoder & Yes & Tests pretrained initialization \\ \addlinespace[0.8pt]
Linear probing & None & Linear readout & Yes & Tests linear accessibility \\ \addlinespace[0.8pt]
Head-only adaptation & None & Nonlinear head or decoder & Yes & Representation plus head capacity \\ \addlinespace[0.8pt]
PEFT & None or restricted & Adapter, LoRA, prompt, bias & Yes & Flexibility--update-cost trade-off \\ \addlinespace[0.8pt]
Direct inference & None & None & No & Requires a compatible output \\ \bottomrule
\end{tabularx}
\end{table*}

\section{Research Landscape and Methods of CFMs}
\label{sec:landscape}

\subsection{Research Landscape and Classification Framework}
\label{sec:classification_framework}

\begin{figure*}[t]
\centering
\includegraphics[width=0.75\textwidth]{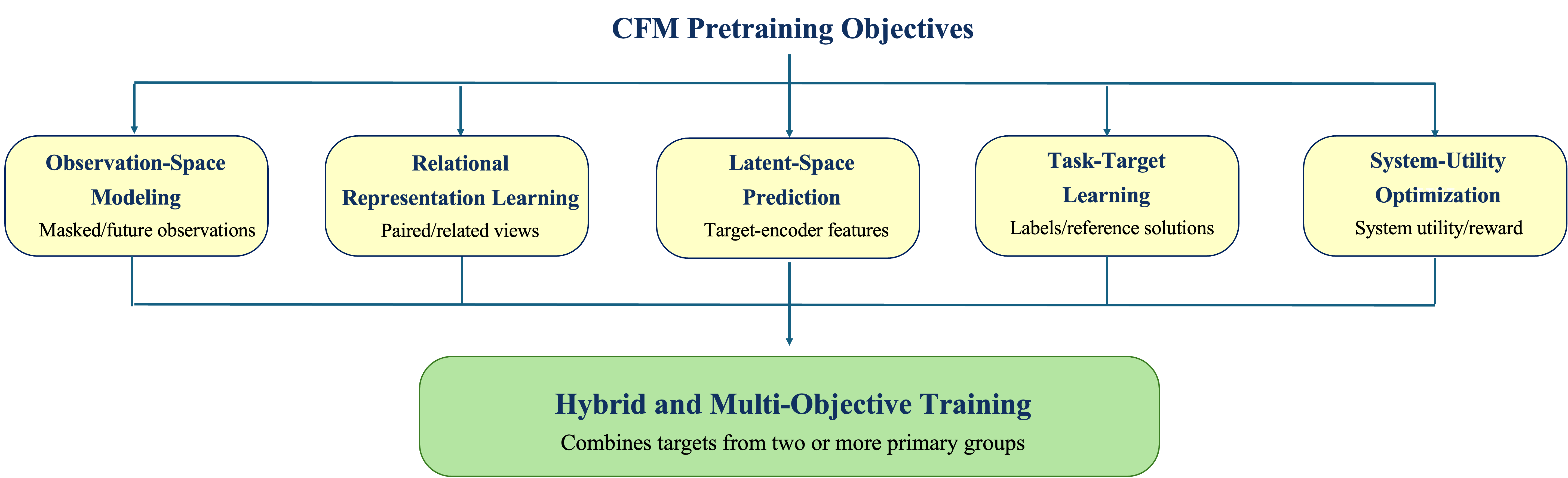}
\caption{Learning-target organization of CFM pretraining methods. The five primary groups are defined by the targets used to update reusable parameters; hybrid and multi-objective training combines targets from two or more primary groups rather than forming an additional primary group.}
\label{fig:landscape}\vspace{-4mm}
\end{figure*}

CFM development connects data, signal representation, backbone design, pretraining, adaptation, application, and inference. Sections~III-B and III-C cover data and backbones; Sections~III-D--III-G address pretraining, adaptation, inference efficiency, and comparison. Fig.~\ref{fig:landscape} organizes pretraining by learning target into observation-space, relational, latent-space, task-target, and system-utility groups. Hybrid methods combine multiple groups. Following Section~II-C, included models retain channel-pretrained parameters for reuse. Input modality, backbone, adaptation, and downstream task are complementary attributes; method tables use objective-appropriate fields within this common organization.

\subsection{Large-Scale Data Resources for CFM Pretraining}
\label{sec:data_resources}

The pretraining data determine the range of environments, radio configurations, mobility patterns, and acquisition conditions that a CFM can learn. Rather than tracing the dataset used by each individual model, Table~\ref{tab:pretraining_data} summarizes large or readily extensible resources for CFM pretraining. It includes measured collections, fixed synthetic and digital-twin datasets, and configurable channel-generation tools, with emphasis on their scale and channel coverage.

CFM pretraining can use CSI, received in-phase/quadrature (IQ) samples, pilot observations, multipath parameters, and synchronized environmental measurements. Reusable datasets should retain the original complex-valued observations, physical units, and acquisition metadata. This information supports frequency-, delay-, angle-, and Doppler-domain transformations without discarding source resolution or calibration.

Measured resources provide complementary forms of radio and environmental diversity. DeepSense~6G~\cite{deepsense6g_dataset} combines millimeter-wave (mmWave) measurements with synchronized visual, ranging, and positioning sensors across numerous scenarios. DICHASUS~\cite{dichasus_dataset} and CAEZ~\cite{wiesmayr2025caez} provide position-tagged massive multiple-input multiple-output (MIMO) and commercial 5G NR measurements, respectively, while WiLoc~\cite{zhang2026wiloc} extends measured Wi-Fi CSI to a substantially larger set of user equipment (UE) locations and access points (APs). WiWorld-RealData~\cite{jiao2026wiworldrealdata} further couples dual-band channel measurements with multiple environmental sensors. These datasets capture hardware and site effects that are difficult to reproduce through simulation, although their channel formats, spatial sampling, and measurement configurations differ.

Synthetic resources offer broader and more controllable coverage. RadioML 2018.01A~\cite{oshea2018overtheair} provides a received-IQ dataset across diverse modulations and channel conditions. CSRD2025~\cite{chang2025csrd2025} describes a reproducible dataset at substantially larger scale and releases its configurable generator rather than directly hosting the approximately 200-TB instance. The WiFo Channel Dataset~\cite{liu2025wifo} sweeps space, time, frequency, and radio configurations with QuaDRiGa, while WAIR-D~\cite{huangfu2022waird} spans city-scale ray-traced environments and SpectrumNet~\cite{zhang2024spectrumnet} covers multiband three-dimensional radio maps. ViWi~\cite{alrabeiah2020viwi}, M$^3$SC~\cite{m3sc_dataset}, SynthSoM~\cite{Cheng2025SynthSoM:AS}, and Multimodal-Wireless~\cite{multimodal_wireless_dataset} associate channel data with virtual-scene observations, while MOCSID~\cite{mocsid_dataset} focuses on spatially consistent multi-cell trajectories. Configurable generation tools provide extensible coverage beyond these released samples. DeepVerse6G~\cite{DeepVerse} generates channel and multimodal observations from indoor and outdoor digital twins, while Raymobtime~\cite{klautau2021raymobtime} adds temporally consistent mobility and synchronized sensing. Geometry-based stochastic tools such as QuaDRiGa~\cite{jaeckel2014quadriga}, NYUSIM~\cite{poddar2024nyusim}, and SEU-PML-6GPCS~\cite{seupml6gpcs,wang2022pervasive6gpcm} complement site-specific ray-tracing platforms such as Sionna RT~\cite{hoydis2023sionnart} and Wireless InSite~\cite{wireless_insite}. DeepMIMO~\cite{alkhateeb2019deepmimo} exposes configurable channels from ray-traced scenarios, while GREAT-X~\cite{huang2025unrealneedmultimodalisac} and LAETwin-XL~\cite{li2026laetwinxl} target multimodal ISAC and low-altitude XL-MIMO settings. Together, these resources support controlled variation of channel and system conditions beyond a fixed released dataset.

\begin{table*}[t]
\caption{Data resources and generation tools for CFM pretraining.}
\label{tab:pretraining_data}
\centering
\footnotesize
\setlength{\tabcolsep}{3.2pt}
\renewcommand{\arraystretch}{1.08}
\begin{tabularx}{\textwidth}{@{}L{3.5cm}L{5.85cm}Y@{}}
\toprule
\textbf{Resource} & \textbf{Data} & \textbf{Scale or coverage} \\ \midrule
\multicolumn{3}{@{}l}{\textit{Measured datasets}} \\ \addlinespace[1pt]
DeepSense 6G~\cite{deepsense6g_dataset} &
mmWave channels; RGB, LiDAR, radar, GNSS &
$>$1 million synchronized samples; $>$30 scenarios \\ \addlinespace[0.8pt]
DICHASUS~\cite{dichasus_dataset} &
Position-tagged massive-MIMO CSI &
Multiple indoor and outdoor campaigns \\ \addlinespace[0.8pt]
CAEZ~\cite{wiesmayr2025caez} &
5G NR uplink CSI from commercial radio units &
4 O-RUs; indoor and outdoor campaigns \\ \addlinespace[0.8pt]
WiLoc~\cite{zhang2026wiloc} &
Wi-Fi CSI; UE locations; AP identifiers &
$>$12 million UE locations; $>$3,000 APs; 16 buildings; $>$30 streets \\ \addlinespace[0.8pt]
WiWorld-RealData~\cite{jiao2026wiworldrealdata} &
Dual-band CIR; cameras, LiDAR, radar, GNSS &
10 TB of measurements; one dual-band route currently released \\ \midrule
\multicolumn{3}{@{}l}{\textit{Synthetic and digital-twin datasets}} \\ \addlinespace[1pt]
RadioML 2018.01A~\cite{oshea2018overtheair} &
Received IQ sequences &
2.56 million sequences; 24 modulations; 26 SNR levels \\ \addlinespace[0.8pt]
CSRD2025~\cite{chang2025csrd2025} &
Generator; IQ frames and annotated spectrograms &
Described: $>$25 million frames; $\sim$200 TB; 100 modulations \\ \addlinespace[0.8pt]
WiFo Channel Dataset~\cite{liu2025wifo} &
QuaDRiGa space--time--frequency CSI &
228,000 samples across 19 radio configurations \\ \addlinespace[0.8pt]
SpectrumNet~\cite{zhang2024spectrumnet} &
3D radio maps; terrain and climate metadata &
5 bands; 11 terrains; 3 climate conditions \\ \addlinespace[0.8pt]
WAIR-D~\cite{huangfu2022waird} &
Channels; user locations; environment maps &
10,000 sparse-user and 100 dense-user environments; $>$40 cities \\ \addlinespace[0.8pt]
ViWi~\cite{alrabeiah2020viwi} &
Ray-traced mmWave channels; RGB images &
Parametric and scalable vision-wireless scenarios \\ \addlinespace[0.8pt]
M$^3$SC~\cite{m3sc_dataset} &
CSI, RGB, depth, LiDAR, radar, and path loss &
200,000 channel matrices; 2.41 TB; synchronized modalities \\ \addlinespace[0.8pt]
SynthSoM~\cite{Cheng2025SynthSoM:AS} &
CSI and multimodal environmental sensing &
$>$9.5 million groups; 1.5 million channel matrices \\ \addlinespace[0.8pt]
Multimodal-Wireless~\cite{multimodal_wireless_dataset} &
CSI, cameras, LiDAR, radar, GNSS, and IMU &
161,400 frames; 4 towns; 16 scenarios; 3 weather conditions \\ \addlinespace[0.8pt]
MOCSID~\cite{mocsid_dataset} &
CSI, CIR, position, velocity, multipath parameters &
17.9 GB; pedestrian trajectories across 10 base stations \\ \midrule
\multicolumn{3}{@{}l}{\textit{Channel and scenario generation tools}} \\ \addlinespace[1pt]
DeepVerse6G~\cite{DeepVerse} &
CSI, RGB, LiDAR, radar, and position data &
Configurable indoor and outdoor digital-twin scenarios \\ \addlinespace[0.8pt]
DeepMIMO~\cite{alkhateeb2019deepmimo} &
Ray-traced MIMO-OFDM channels &
Configurable arrays, users, bandwidths, and subcarriers \\ \addlinespace[0.8pt]
Raymobtime~\cite{klautau2021raymobtime} &
MIMO channels; RGB, LiDAR, and mobility &
Configurable temporally consistent mobile scenarios \\ \addlinespace[0.8pt]
QuaDRiGa~\cite{jaeckel2014quadriga} &
Geometry-based stochastic channels &
Multi-link, mobility, array, frequency, and scenario sweeps \\ \addlinespace[0.8pt]
NYUSIM~\cite{poddar2024nyusim} &
Statistical spatial channels &
0.5--150 GHz; urban, rural, indoor hotspot, and factory scenarios \\ \addlinespace[0.8pt]
Sionna RT~\cite{hoydis2023sionnart} &
Differentiable site-specific ray tracing &
Configurable scenes, materials, arrays, frequencies, and mobility \\ \addlinespace[0.8pt]
Wireless InSite~\cite{wireless_insite} &
Commercial site-specific ray tracing &
Geometry, materials, antennas, propagation mechanisms \\ \addlinespace[0.8pt]
SEU-PML-6GPCS~\cite{seupml6gpcs,wang2022pervasive6gpcm} &
6G pervasive geometry-based stochastic channels &
Sub-6 GHz--optical; terrestrial, aerial, satellite, maritime, RIS, ISAC \\ \addlinespace[0.8pt]
GREAT-X~\cite{huang2025unrealneedmultimodalisac} &
Unreal Engine CSI, RGB, radar, LiDAR &
Configurable multimodal ISAC scenarios \\ \addlinespace[0.8pt]
LAETwin-XL~\cite{li2026laetwinxl} &
Low-altitude XL-MIMO ray-traced channels &
Configurable aerial trajectories and array geometries \\ \bottomrule
\end{tabularx}
\vspace{1pt}
\parbox{\textwidth}{\scriptsize\textit{Abbreviations:} CIR, channel impulse response; O-RU, Open RAN radio unit; GNSS, global navigation satellite system; IMU, inertial measurement unit; OFDM, orthogonal frequency-division multiplexing; RGB, red--green--blue; LiDAR, light detection and ranging; SNR, signal-to-noise ratio.}\vspace{-4mm}
\end{table*}

No single resource spans the full range of channel conditions required for a general CFM. Measured datasets capture radio hardware and environmental effects but are costly to scale, whereas synthetic data provide controlled coverage while inheriting assumptions from their propagation engines. A practical pretraining collection can combine both sources, with source-aware sampling and data partitions that separate locations, trajectories, sessions, devices, and generation engines where appropriate. Reported sample counts are more informative when the counting unit is specified, because a channel snapshot, trajectory, antenna--frequency patch, IQ segment, and synchronized multimodal frame are not equivalent training examples.

\subsection{Generic Backbones and Channel-Structure-Aware Designs}
\label{sec:backbones}

The backbone and pretraining objective address complementary design questions. The objective specifies the learning signal used to construct the base representation, whereas the backbone determines how tokens are mixed and transformed; tokenization and positional encoding connect the observations to that backbone. Transformers are common because attention can couple distant elements across time, frequency, space, and antennas, while patch-token vision Transformers (ViTs) and shifted-window attention provide widely adopted generic antecedents~\cite{dosovitskiy2021vit,liu2021swin}. Most CFM encoders nevertheless retain generic Transformer, convolutional neural network (CNN), or sequence-model blocks. Training a generic backbone on CSI may adapt its learned parameters to channel data, but does not by itself introduce channel-specific inductive biases into the architecture.

Channel-specific inductive biases can be incorporated at different stages of model design. Such biases may be encoded in the input representation, tokenization or positional encoding, the backbone operator, the training objective, or a frontend attached to a generic encoder. Multiple mechanisms may be combined, and this architectural characterization is distinct from the taxonomy of pretraining objectives.

AirFM-DDA~\cite{bian2026airfmdda} combines a delay--Doppler--angle (DDA) representation, local-window attention, and frame-structure-aware positional encoding. Adaptive 3D-RoPE~\cite{zhang2026adaptive3drope} extends the rotary-position principle~\cite{su2024roformer} within a ViT encoder--decoder by replacing a static one-dimensional positional prior with axis-decoupled three-dimensional rotary frequencies modulated by CSI-level statistics.

Explicit modifications to the mixing operator are less common. ConsisFormer~\cite{wang2026consisformer} introduces adaptive token aggregation, equivalent-position tracking, and feature supplementation based on local channel consistency. LWM-Temporal~\cite{alikhani2026lwmtemporal} restricts attention to local angle--delay neighborhoods and feasible temporal corridors, coupling an angle--delay--time representation with sparse spatio-temporal attention. Full-Domain Coupler~\cite{chen2026fulldomaincoupler} interleaves complex-valued mixing operators across time, space, and frequency.

Alternative computational mechanisms require the same distinction. WiMamba~\cite{raviv2026wimamba} builds on the Mamba selective state-space architecture~\cite{gu2024mamba}; SpikeWFM~\cite{jing2026spikewfm} combines a spiking neural network (SNN)~\cite{maass1997spiking} with conventional artificial neural network blocks; and ComHymba~\cite{yang2026comhymba} combines window attention with state-space blocks. These architectures offer alternative efficiency or robustness trade-offs, but the operator family alone does not constitute a channel-specific inductive bias. In ComHymba, for example, channel structure is incorporated through three-dimensional patching, structured masks, and complex amplitude--phase reconstruction losses. phys-WFM~\cite{wang2026physwfm} augments a generic masked autoencoder (MAE) with phase-ramp alignment and recovery to impose time-, frequency-, and space-domain wave equivariance. Filter-and-Attend~\cite{wang2025filterattend} places a noise-plus-interference (NPI) suppression module before a frozen Transformer.

Overall, most existing CFMs retain generic Transformer, state-space, or neural-network blocks and introduce channel structure through the signal representation, tokenization, positional encoding, structured masking, or an auxiliary frontend. Only a smaller group modifies the backbone operator itself. Channel-structure-aware backbones can model multidimensional channel dependencies and complement the choice of pretraining objective.

\subsection{CFM Pretraining Objectives}
\label{sec:pretraining_objectives}

Pretraining methods are organized according to the form of their learning targets: channel or signal observations, relations between representations, predicted latent features, task outputs, or communication-system utilities. Hybrid and multi-objective methods combine two or more of these targets when optimizing the shared pretrained parameters. The following subsections review the learning mechanisms and representative CFM methods in each group. In the accompanying method tables, the \emph{Open} column indicates the availability of an implementation of the listed pretraining method or its pretrained weights: ``Both'' denotes the release of both, ``Code'' and ``Weights'' denote the corresponding individual release, and ``/'' indicates that no official code or pretrained weights were identified at the review cutoff.

\subsubsection{Observation-Space Modeling}
\label{sec:observation_space}

Observation-space pretraining optimizes targets derived from the input observations and expressed in an observable channel or signal space, including CSI, IQ, spectra, and deterministic transforms of these observations. It encompasses reconstruction and denoising, including masked modeling, as well as direct or autoregressive observation prediction and diffusion-based completion or generation.

\begin{figure*}[t]
\centering
\begin{minipage}[t]{0.49\textwidth}
\centering
\includegraphics[width=\linewidth]{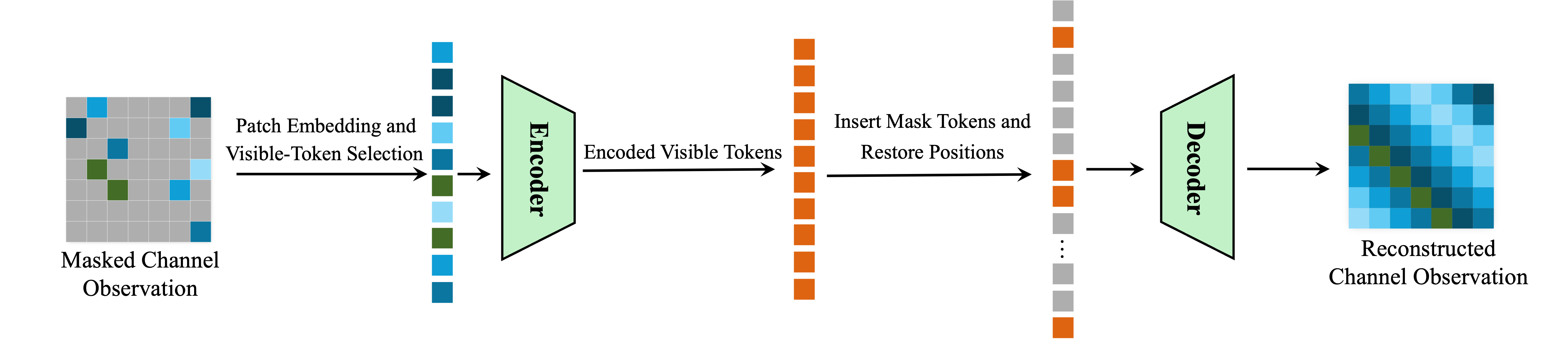}\\[-1pt]
\footnotesize (a) Observation-space masked reconstruction
\end{minipage}\hfill
\begin{minipage}[t]{0.49\textwidth}
\centering
\includegraphics[width=\linewidth]{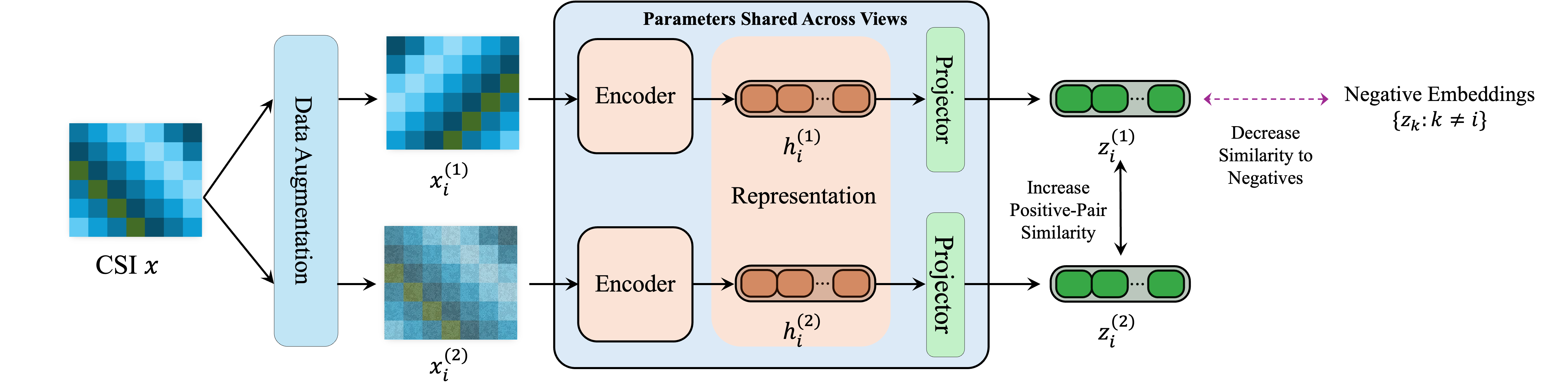}\\[-1pt]
\footnotesize (b) Relational representation learning
\end{minipage}
\caption{Two major CFM pretraining mechanisms. (a) The encoder processes visible channel tokens and the decoder reconstructs withheld observations. (b) Shared encoders map related channel views to representations whose relation is optimized through contrastive or related objectives.}
\label{fig:pretraining_mechanisms}\vspace{-4mm}
\end{figure*}

Masked channel modeling is a prominent example. Fig.~\ref{fig:pretraining_mechanisms}(a) shows a patch-based formulation: an encoder processes visible tokens, mask tokens restore the sequence layout, and a decoder reconstructs the withheld observations. LWM~\cite{alikhani2024large}, BERT4MIMO~\cite{catak2025bert4mimo}, and MCM~\cite{guo2026mcm} apply this design to CSI; WiFo~\cite{liu2025wifo} adds random and axis-specific masks, while 6G WavesFM~\cite{aboulfotouh20256g} and multimodal variants extend it to spectrogram or IQ inputs~\cite{aboulfotouh2025multimodalwireless}. Reconstruction losses may be evaluated in the original observation space or a deterministic signal-domain transform, and masking may operate on samples, coefficients, patches, or tokens.

Mask geometry influences the conditional structure that the model is trained to use. Uniform random masking primarily encourages generic completion, whereas structured masks can emulate sparse sounding, forecasting, or extrapolation. LWM-Temporal~\cite{alikhani2026lwmtemporal} uses a physics-informed curriculum of random, rectangular, pilot-pattern, and spatio-temporal masks in the angle--delay--time domain; ComHymba~\cite{yang2026comhymba} combines dimension-specific, block, and pilot-pattern masks with three-dimensional patching. ComHymba further applies a weighted amplitude--phase loss to account separately for magnitude and phase reconstruction errors. These designs illustrate that the learned representation depends not only on which observations are masked, but also on how reconstruction fidelity is defined.

Observation-space prediction also includes future samples, unobserved frequencies or antennas, and the next observation token in a context sequence. Y. Sheng et al.~\cite{sheng2025wirelessfoundationmodelmultitask}, for example, directly extrapolate channel sequences and predict ISAC AoA and traffic sequences through multi-step regression. Autoregressive token prediction provides an alternative formulation of these prediction problems.

Diffusion provides a probabilistic alternative to direct reconstruction and forecasting. Denoising diffusion and score-based models learn iterative reverse processes from corrupted observations~\cite{ho2020ddpm,song2021score}. LAETwin-XL~\cite{li2026laetwinxl} conditions such a process on incomplete channel observations to generate complete channel realizations. Unlike deterministic predictors that produce a single estimate, conditional diffusion can represent multiple channel realizations consistent with the same partial observation, which is useful when the missing channel components are not uniquely determined.

Table~\ref{tab:observation_methods} compares the pretraining observations, designs, and downstream tasks of representative methods in this group.

\begin{sidewaystable*}[p]
\caption{Representative observation-space modeling methods grouped by prediction form. Reported percentages denote mask ratios.}
\label{tab:observation_methods}
\centering
\footnotesize
\setlength{\tabcolsep}{2.8pt}
\renewcommand{\arraystretch}{1.05}
\resizebox{0.98\textheight}{!}{%
\begin{tabular}{@{}L{2.9cm}L{4.3cm}L{4.2cm}L{10.9cm}C{1.3cm}@{}}
\toprule
\textbf{Model} & \textbf{Pretraining data} & \textbf{Objective or design} & \textbf{Downstream evaluations} & \textbf{Open} \\ \midrule
\multicolumn{5}{@{}l}{\textbf{\textit{Reconstruction and denoising}}} \\ \addlinespace[0.8pt]
Aboulfotouh et al.~\cite{aboulfotouh2024building} & Spectrograms & Masked spectrogram modeling (70--80\%) & Human Activity Recognition; NR and LTE Spectrogram Segmentation & / \\
6G WavesFM~\cite{aboulfotouh20256g} & CSI and spectrograms & Masked wireless modeling (75\%) & 5G NR Positioning; MIMO-OFDM Channel Estimation; RF Signal Classification; Human Activity Recognition & / \\
Aboulfotouh et al.~\cite{aboulfotouh2025multimodalwireless} & IQ samples and spectrograms & Multimodal masked wireless modeling (70\%) & Human Activity Recognition; RF Signal Classification; 5G NR Positioning; RF Device Fingerprinting; Interference Detection; Interference Classification & Both \\
Adaptive 3D-RoPE~\cite{zhang2026adaptive3drope} & Space--time--frequency CSI & Adaptive 3D rotary encoding with random and axis-specific masking & CSI Reconstruction; Time- and Frequency-Domain Channel Extrapolation & Code \\
AirFM-DDA~\cite{bian2026airfmdda} & Delay--Doppler--angle CSI & Task-structured masked DDA reconstruction & Time- and Frequency-Domain Channel Extrapolation; Channel Estimation & / \\
Cheraghinia et al.~\cite{cheraghinia2025unifiedwireless} & IQ and CIR sequences & Random patch reconstruction & Short-/Long-Range Technology Recognition; LoS/NLoS Classification; UWB Ranging Correction & / \\
ComHymba~\cite{yang2026comhymba} & Space--time--frequency CSI & Domain-informed structured masking with amplitude--phase reconstruction & Channel Estimation; Channel Extrapolation; FDD Inference; Positioning; Path-Loss Prediction; LoS/NLoS Classification; Scenario Classification; Cross-Frequency Beam Prediction & / \\
BERT4MIMO~\cite{catak2025bert4mimo} & CSI & Masked CSI reconstruction & CSI Reconstruction in Held-Out Scenarios & Code \\
Jiang et al.~\cite{jiang2025scalewificfm} & Wi-Fi CSI from multiple datasets & Random masking (80\%) & Human Activity Recognition; Gesture Recognition; User Identification & / \\
CSI-MAE~\cite{jiang2026csimaemaskedautoencoderbasedchannel} & Antenna--frequency CSI & Random masking (75\%) & Antenna- and Frequency-Domain Channel Extrapolation; CSI Feedback; Positioning & Code \\
FM-RME~\cite{yang2026fmrme} & Spatial--temporal--spectral radio maps & Masked multidimensional radio-map estimation & Spatial, Temporal, and Spectral Radio-Map Estimation & / \\
HeterCSI~\cite{zhang2026hetercsi} & Variable-size CSI & MAE with padding-aware attention & CSI Reconstruction; Time- and Frequency-Domain Channel Extrapolation & Both \\
Chen et al.~\cite{chen2026willm} & Heterogeneous Wi-Fi CSI amplitudes & Dataset-specific tokenizers and reconstruction heads & Human Activity and Gesture Recognition; Human Identification; Fall Detection; Crowd Counting & Code \\
ConsisFormer~\cite{wang2026consisformer} & Antenna--frequency CSI & Consistency-driven aggregated autoencoding & Frequency-Domain Channel Extrapolation; LoS/NLoS Classification; Beam Prediction; Positioning & / \\
Filter-and-Attend~\cite{wang2025filterattend} & Received pilot and payload signals & Supervised NPI suppression and CSI reconstruction & Time- and Frequency-Domain Channel Extrapolation; Positioning & / \\
LWM~\cite{alikhani2024large} & Antenna--frequency CSI & Masked channel modeling (15\%) & Sub-6-to-mmWave Beam Prediction; LoS/NLoS Classification & Both \\
LWM-Spectro~\cite{kim2026lwmspectro} & Wi-Fi, LTE, and 5G spectrograms & Masked spectrogram modeling (70\%) & Modulation Classification; Joint SNR/Doppler Classification & Weights \\
LWM-Temporal~\cite{alikhani2026lwmtemporal} & Angle--delay--time CSI & Physics-informed masking curriculum & Time-Domain Channel Extrapolation & Both \\
MCM~\cite{guo2026mcm} & CSI & Masked channel modeling & Time-Domain Channel Extrapolation; Channel Estimation; CSI Feedback & / \\
PilotWiMAE~\cite{guler2026pilotwimae} & Noisy pilot observations & Factorized random masking (99\%, then 79\%) & Cross-Frequency Beam Prediction; LoS/NLoS Classification; Channel Estimation & Both \\
Ott et al.~\cite{ott2024radiofoundation} & Measured CIR & Masked CIR reconstruction & Fingerprint-Based Indoor Positioning & / \\
phys-WFM~\cite{wang2026physwfm} & Heterogeneous space--time--frequency CSI & Wave-equivariant random and axis-specific masking & CSI Reconstruction; Time- and Frequency-Domain Channel Extrapolation & / \\
SpikeWFM~\cite{jing2026spikewfm} & Space--time--frequency CSI & Masked CSI reconstruction (75\%) & Time-Domain Channel Extrapolation & / \\
WALoMA~\cite{makin2026waloma} & CSI & Random patch masking (75\%) & LoS/NLoS Classification; Beam Prediction; Channel Interpolation; Channel Estimation; Channel Charting & / \\
WiFo-CF~\cite{liu2026wifocf} & Variable-size CSI & Multiuser, multirate reconstruction & CSI Feedback; Positioning & / \\
WiFo~\cite{liu2025wifo} & Space--time--frequency CSI & Random and axis-specific masking & Time- and Frequency-Domain Channel Extrapolation & Both \\
WiFo-2~\cite{liu2025wifo2} & Space--time--frequency CSI & Masked reconstruction, denoising, and confidence prediction & Channel Estimation; Channel Extrapolation (Time-/Frequency-Domain, Cross-Band, Vision-Aided); LoS/NLoS Classification; Cross-Band Beam Prediction; Positioning; AoA Estimation; Doppler Estimation; CSI Feedback; Signal Detection & Both \\
WiMamba~\cite{raviv2026wimamba} & Antenna--frequency CSI & Masked token reconstruction & LoS/NLoS Classification; Beam Prediction; Channel Interpolation; Positioning & Code \\
WirelessGPT~\cite{yang2025wirelessgpt} & Space--time--frequency CSI & Random patch reconstruction & Channel Estimation; Channel Extrapolation; Human Activity Recognition; Environment Reconstruction & / \\
Zhang et al.~\cite{zhang2026crossbandcsi} & Sparse noisy single-band pilots & Pilot-densification reconstruction & Cross-Band CSI Reconstruction from Source-Band CSI and Target-Band Pilots & / \\ \midrule
\multicolumn{5}{@{}l}{\textbf{\textit{Direct prediction}}} \\ \addlinespace[0.8pt]
Y. Sheng et al.~\cite{sheng2025wirelessfoundationmodelmultitask} & Channel, AoA, and traffic sequences & Direct multi-step forecasting with patch masking & Channel Extrapolation; ISAC AoA Prediction; Traffic Prediction; ISAC Time-Delay Prediction & / \\ \midrule
\multicolumn{5}{@{}l}{\textbf{\textit{Diffusion, flow, and score-based modeling}}} \\ \addlinespace[0.8pt]
FARM~\cite{gao2026farm} & Aerial radio maps and environmental priors & Masked denoising with flow matching (75\%) & Condition-Free and Condition-Based Aerial Radio-Map Construction & / \\
LAETwin-XL~\cite{li2026laetwinxl} & Partially observed XL-MIMO CSI & Conditional denoising diffusion implicit model (DDIM) & Channel Estimation; Channel Extrapolation; Near-/Far-Field Classification; 3D Positioning & / \\
ScoreFM~\cite{cai2026scorefm} & Channels, source signals, and structured interference & Denoising score matching & Channel Estimation; Interference-Aware Positioning; Blind Semantic Communication & / \\ \bottomrule
\end{tabular}
}
\end{sidewaystable*}

Masked reconstruction is the dominant observation-space objective among existing methods. Existing approaches differ primarily in how the observable context and prediction target are defined, which determines the channel dependencies emphasized during pretraining. Direct prediction focuses on a specified target, whereas diffusion and score-based methods model conditional uncertainty. Downstream behavior is jointly shaped by the pretraining data, backbone, decoder, and adaptation protocol.

\subsubsection{Relational Representation Learning}
\label{sec:contrastive}

Relational objectives specify which representations are aligned and which are separated. Fig.~\ref{fig:pretraining_mechanisms}(b) shows a common construction in which shared encoders and projectors process two related channel views. Contrastive implementations obtain negative embeddings through in-batch sampling or momentum dictionaries and may also align paired modalities~\cite{chen2020simple,he2020moco,radford2021clip}. Non-contrastive and self-distillation methods instead use target encoders or representation regularization, while triplet losses impose an explicit margin. All learn relations among channel representations rather than reconstructing observations.

Positive pairs can be augmentations of one observation, synchronized modalities, related channel views, or measurements linked by metadata. CSI-CLIP~\cite{jiang2025mimo} and CSI-CLIP++~\cite{jiang2026csiclippp} align frequency-domain CSI with delay-domain CIR; CSI-CLIP++ retains this objective while scaling the backbone and evaluation beyond CSI-CLIP. IQFM~\cite{mashaal2025iqfm} and the method of O. Kanu et al.~\cite{kanu2025selfsupervisedradiorepresentationlearning} instead construct augmented pairs of received signals. SWiT~\cite{salihu2024selfsupervised} matches augmented global and local CSI views through a momentum target encoder without negative pairs, and transfers the learned representation across environments and to positioning and path-loss prediction. In the paired CSI--CIR setting, the two inputs are deterministic Fourier-related views of the same channel realization. Such shared information can facilitate alignment, although it may also introduce shortcuts that do not require separation of individual multipath variables. Augmentation design and negative-sample selection are therefore particularly important because physically similar channels can otherwise be treated as unrelated samples.

For a minibatch of paired embeddings $\{(\mathbf{z}_i,\mathbf{z}_i^+)\}_{i=1}^{N}$, a common InfoNCE form is

\begin{small}
\begin{equation}
\mathcal{L}_{\mathrm{InfoNCE}} =
-\frac{1}{N}\sum_{i=1}^{N}
\log
\frac{\exp(\operatorname{sim}(\mathbf{z}_i,\mathbf{z}_i^+)/\tau)}
{\sum_{j=1}^{N}\exp(\operatorname{sim}(\mathbf{z}_i,\mathbf{z}_j^+)/\tau)},
\end{equation}\end{small}
where $\operatorname{sim}(\cdot,\cdot)$ is typically cosine similarity and $\tau$ is a temperature. Triplet objectives implement the same relational principle by requiring an anchor to be closer to a positive than to a negative by a specified margin.

Traditional channel charting provides an important antecedent to relational channel representation learning. It constructs a geometry-preserving low-dimensional embedding, often for a particular cell or environment, using triplet, autoencoding, or manifold-learning techniques~\cite{8444621,ferrand2021triplet,ferrand2023wireless}. In CFM research, channel charting can provide either a pretraining signal or a downstream application. Reuse beyond the original environment depends on the associated pretraining and transfer setting.

CSI2Vec~\cite{palhares2025csi2vec} trains a shared self-supervised encoder over heterogeneous CSI datasets using a triplet objective and reuses the resulting representation for positioning and channel charting. Its downstream heads are scenario-specific, and the reported evaluation provides evidence of cross-task reuse within the domains observed during encoder training.

Table~\ref{tab:relational_methods} compares the related views, learning objectives, and downstream tasks.

\begin{table*}[t]
\caption{Representative relational representation-learning methods.}
\label{tab:relational_methods}
\centering
\scriptsize
\setlength{\tabcolsep}{1.0pt}
\renewcommand{\arraystretch}{1.08}
\begin{tabularx}{\textwidth}{@{}L{2.0cm}L{2.55cm}L{2.9cm}Y C{0.8cm}@{}}
\toprule
\textbf{Model} & \textbf{Related views} & \textbf{Objective} & \textbf{Downstream tasks} & \textbf{Open} \\ \midrule
CSI-CLIP~\cite{jiang2025mimo} & CSI--CIR pairs & Cross-view contrast & LoS/NLoS Classification; Beam Prediction; Positioning & Both \\
CSI-CLIP++~\cite{jiang2026csiclippp} & CSI--CIR pairs & Cross-view contrast & LoS/NLoS Classification; Beam Prediction; Positioning & / \\
CSI2Vec~\cite{palhares2025csi2vec} & Temporal CSI triplets & Triplet metric learning & Positioning; Channel Charting & / \\
IQFM~\cite{mashaal2025iqfm} & Augmented IQ pairs & Task-aware contrast & Modulation Classification; AoA Classification; Beam Prediction; RF Device Fingerprinting & / \\
Kanu et al.~\cite{kanu2025selfsupervisedradiorepresentationlearning} & Augmented signal pairs & Momentum contrast & AoA Estimation; Modulation Classification & / \\
SWiT~\cite{salihu2024selfsupervised} & Global/local CSI views & Multi-view self-distillation & Positioning; Path-Loss Prediction & Both \\
WMFM~\cite{farzanullah2025wmfm} & CSI--RGB pairs & Cross-modal contrast & Positioning; LoS/NLoS Classification & / \\
WiFo-M$^2$~\cite{zhang2026wifom2} & Sensing--CSI sequences & Cross-modal contrast & Beam Prediction; Channel Estimation, Interpolation, and Extrapolation; Positioning & Weights \\ \bottomrule
\end{tabularx}\vspace{-4mm}
\end{table*}

Most methods use contrastive alignment, whereas CSI2Vec adopts triplet-based metric learning and SWiT uses non-contrastive self-distillation. Across these approaches, the learned invariances depend strongly on how relations between observations are defined. Relational objectives avoid reconstructing every observation detail, but they may suppress task-relevant channel variations or exploit shortcuts when the training relations are not physically meaningful. Their transferability should therefore be evaluated under downstream tasks and channel conditions that were not used to construct the pretraining relations.

\subsubsection{Latent-Space Prediction}
\label{sec:jepa}

\begin{figure}[!t]
\centering
\includegraphics[width=0.75\columnwidth]{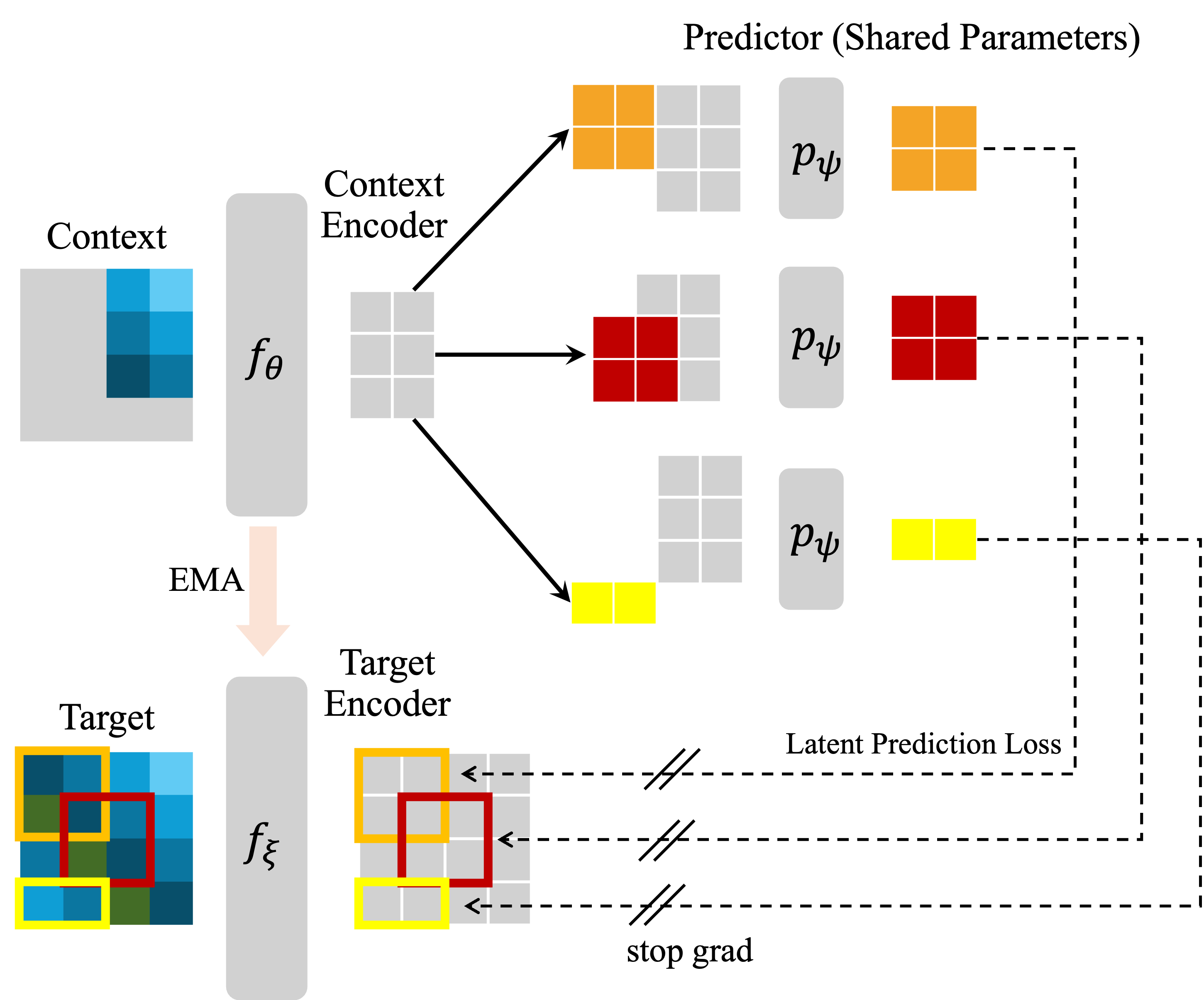}
\caption{Multi-target JEPA-style latent prediction. A context encoder and shared predictor estimate stop-gradient target-encoder representations of multiple withheld regions; the target encoder is updated using an EMA.}
\label{fig:jepa_pipeline}\vspace{-4mm}
\end{figure}

Latent predictive methods replace an observation-space target with a representation produced from a distinct target region or future state. Fig.~\ref{fig:jepa_pipeline} illustrates a multi-target joint-embedding predictive architecture (JEPA)~\cite{assran2023selfsupervised}. A context encoder represents the visible context, and a predictor with parameters shared across target regions combines this representation with region-specific target-location conditioning. For target regions $r\in\mathcal{R}$, a generic formulation is

\begin{small}
\begin{equation}
\begin{aligned}
\mathbf{z}_{t,r}
&=\operatorname{sg}\!\left(S_r\!\left(f_{\boldsymbol{\xi}}(\mathbf{x}_{t})\right)\right),\\
\hat{\mathbf{z}}_{t,r}
&=p_{\boldsymbol{\psi}}\!\left(f_{\boldsymbol{\theta}}(\mathbf{x}_{c}),\mathbf{c}_{r}\right),\\
\mathcal{L}_{\mathrm{JEPA}}
&=\frac{1}{|\mathcal{R}|}\sum_{r\in\mathcal{R}}
d(\hat{\mathbf{z}}_{t,r},\mathbf{z}_{t,r}),
\end{aligned}
\end{equation}\end{small}where $\mathbf{x}_{c}$ and $\mathbf{x}_{t}$ are the context and target observations, respectively; $S_r$ selects target region $r$ from the encoded target; and $\mathbf{c}_{r}$ supplies its location or other optional conditioning. The stop-gradient operator $\operatorname{sg}$ prevents direct gradient flow through the target representations. As depicted in the figure, the target encoder may be updated from the context encoder using an exponential moving average (EMA); other mechanisms may also be used to avoid representational collapse.

WirelessJEPA~\cite{chu2026wirelessjepa} applies spatio-temporal latent prediction to multi-antenna IQ observations, while LatentWave~\cite{mohamed2026latentwave} extends JEPA-style pretraining across CSI and spectrogram inputs. CSI-JEPA~\cite{luo2026csijepa} and WiFi-JEPA~\cite{kim2026wifijepa} learn predictive CSI representations for multitask Wi-Fi sensing and cross-environment 3D pose estimation, respectively. JEPA-MSAC~\cite{zheng2026jepamsac} maps vision, radar, LiDAR, GPS, and RF observations into a shared predictive latent space, whereas WWM~\cite{chen2026wirelessworldmodel} combines CSI with environmental point clouds and trajectory context. These methods predict target-encoder representations of masked regions or modalities; structured masks and multimodal context determine which spatial, temporal, spectral, or cross-modal relations must be inferred.

Table~\ref{tab:latent_methods} compares the context inputs, latent targets, and downstream tasks.

\begin{table*}[t]
\caption{Representative latent-space predictive methods.}
\label{tab:latent_methods}
\centering
\small
\setlength{\tabcolsep}{2.5pt}
\renewcommand{\arraystretch}{1.15}
\begin{tabularx}{\textwidth}{|L{2.6cm}|L{3.0cm}|L{3.0cm}|Y|C{0.75cm}|}
\hline
\textbf{Model} & \textbf{Context} & \textbf{Latent target} & \textbf{Downstream tasks} & \textbf{Open} \\ \hline
CSI-JEPA~\cite{luo2026csijepa} & Wi-Fi CSI amplitudes & Masked-region latents & Fall and Breathing Detection; Motion-Source, Human-Activity, and Proximity Recognition; User Identification; \mbox{Room-Level} Positioning & / \\ \hline
JEPA-MSAC~\cite{zheng2026jepamsac} & Vision, radar, LiDAR, GPS, and RF & Temporal-block latents & Positioning; Beam Prediction; Received Signal Strength Indicator (RSSI) Prediction & / \\ \hline
LatentWave~\cite{mohamed2026latentwave} & CSI and spectrograms & Masked-region latents & RF Signal Classification; 5G NR Positioning; Beam Prediction; LoS/NLoS Classification & / \\ \hline
WiFi-JEPA~\cite{kim2026wifijepa} & Real/simulated \mbox{multi-link} CSI & Masked-link latents & Single- and Multi-Person 3D Pose Estimation & / \\ \hline
WirelessJEPA~\cite{chu2026wirelessjepa} & Antenna--time IQ & Masked-region latents & AoA Estimation; RF Device Fingerprinting; Modulation Classification; Protocol Classification; GNSS Jamming Classification; Interference Classification & / \\ \hline
WWM~\cite{chen2026wirelessworldmodel} & CSI, point clouds, and trajectories & CSI/trajectory latents & Time- and Frequency-Domain Channel Extrapolation; CSI Feedback; Beam Prediction; Positioning & / \\ \hline
\end{tabularx}
\end{table*}

Latent-space prediction differs from view-level self-distillation because the context encoder predicts the target representation of withheld or future observations rather than aligning complete views. This formulation avoids reconstructing every channel coefficient, but the learned target remains a latent feature whose direct physical interpretation is not guaranteed. Its value should therefore be assessed through downstream transfer and controlled tests of the channel information retained in the learned representation.

\subsubsection{Task-Target Learning}
\label{sec:task_target}

Task-target learning uses outputs defined by downstream problems, including ground-truth annotations, teacher predictions, and algorithm-generated reference solutions. Unlike observation-space modeling, whose targets are constructed from withheld or future channel observations, task-target objectives are defined by the semantics of downstream outputs. Typical formulations include regression losses for channel quantities or positions, cross-entropy losses for discrete decisions, and imitation losses for precoders, schedules, or resource allocations. The UBERT variant of BERT4beam~\cite{li2025bert4beam} learns from optimization-generated beamformers across utility functions and antenna scales. FM-Receiver~\cite{zheng2026fmreceiver} jointly supervises transmitted bits, channel coefficients, and multiplicative noise in a unified inner/outer receiver. WiFo-MUD~\cite{yang2026wifomud} distills a heterogeneous multi-user detector for transfer across unseen channels and system configurations. OmniLoc~\cite{chu2026omniloc} jointly predicts the building, floor, and UE position and evaluates reuse through zero-shot inference, LoRA, and frozen-backbone head tuning across environments. LocGPT~\cite{zhao2024locgpt} and GLocFM~\cite{bian2026glocfm} reuse positioning backbones across scenarios and across scenes with simulation-to-real transfer, respectively.

Table~\ref{tab:task_target_methods} compares the task targets and reuse settings of these methods.

\begin{table*}[t]
\caption{Representative task-target pretraining methods.}
\label{tab:task_target_methods}
\centering
\small
\setlength{\tabcolsep}{2.5pt}
\renewcommand{\arraystretch}{1.10}
\begin{tabularx}{\textwidth}{@{}L{2.6cm}L{4.2cm}L{3.1cm}Y C{0.75cm}@{}}
\toprule
\textbf{Model} & \textbf{Task targets} & \textbf{Reused component} & \textbf{Reuse setting} & \textbf{Open} \\ \midrule
UBERT~\cite{li2025bert4beam} & Optimized Beamformers & UBERT Backbone & Cross-Utility and Cross-Scale Beamforming & / \\
FM-Receiver~\cite{zheng2026fmreceiver} & Bits; CSI; Multiplicative Noise & Inner/Outer Receiver & Band/Channel-Model/Configuration Transfer & / \\
WiFo-MUD~\cite{yang2026wifomud} & Transmitted Symbols & Distilled WiFo-MUD & Unseen Channels and System Configurations & / \\
OmniLoc~\cite{chu2026omniloc} & Building; Floor; UE Position & OmniLoc Backbone & Cross-Environment Positioning & Both \\
LocGPT~\cite{zhao2024locgpt} & UE Position & LocGPT Backbone & Cross-Scenario Positioning & Code \\
GLocFM~\cite{bian2026glocfm} & 3D Tx Position & GLocFM Backbone & Cross-Scene and Sim-to-Real Positioning & / \\ \bottomrule
\end{tabularx}\vspace{-4mm}
\end{table*}

Task-target supervision directly aligns the learned representation with predefined outputs, although its transferability depends on whether the retained shared parameters remain useful beyond the tasks, labels, and reference-solution procedures used during training. Current studies mainly demonstrate reuse across a predefined task family or across operating configurations and channel models.

\subsubsection{System-Utility Optimization}
\label{sec:system_utility}

System-utility optimization maps the model output to a precoder, schedule, power allocation, or other system variable and evaluates it through a communication-system utility rather than a reference action. When the path is differentiable, a generic form is

\begin{small}
\begin{equation}
    \widehat{\mathbf{d}}
    =g_{\boldsymbol{\phi}}\!\left(
    f_{\boldsymbol{\theta}}(\mathbf{x}),\mathbf{c}\right),
    \qquad
    \mathcal{L}_{\mathrm{util}}
    =-\mathbb{E}_{\mathbf{s}}
    \left[U(\widehat{\mathbf{d}};\mathbf{s})\right],
\end{equation}\end{small}where $\widehat{\mathbf{d}}$ is the model-generated system output, $\mathbf{c}$ is optional context, $\mathbf{s}$ contains the channel and system state, and $U$ can encode rate, energy efficiency, or a quality-of-service (QoS) criterion. Non-differentiable simulators or delayed system feedback can instead supply a reward optimized by policy-gradient or other direct-optimization procedures. These objectives avoid reference action labels but still require channel observations, simulated system states, or operational feedback.

AWF~\cite{tong2026awf} maximizes worst-case mutual information in adversarial water-filling and reuses its frozen learned primal--dual model across unseen problem sizes, constraint families, and modulation formats. WiFo-E~\cite{wen2026wifoe} maximizes expected sum rate across heterogeneous frequency-division-duplexing configurations and reuses its shared trunk when adapting new configuration-specific branches. J. Emery et al.~\cite{emery2025massivemimoprecoding} use sequential utility-oriented phases while reusing a shared feature extractor across deployment sites. SiFo~\cite{zhao2026sifo} directly maximizes CSI-capture efficiency during cross-site pretraining and reuses its synchronization signal block (SSB) codebook and subspace predictor for target-site CSI feedback after calibration.

Table~\ref{tab:system_utility_methods} compares the utility targets, reused components, and reuse settings of these methods.

\begin{table*}[t]
\caption{Representative system-utility CFM pretraining methods.}
\label{tab:system_utility_methods}
\centering
\small
\setlength{\tabcolsep}{2.2pt}
\renewcommand{\arraystretch}{1.10}
\begin{tabularx}{\textwidth}{@{}L{2.6cm}L{4.2cm}L{5.0cm}Y C{0.75cm}@{}}
\toprule
\textbf{Model} & \textbf{System utility} & \textbf{Reused component} & \textbf{Reuse setting} & \textbf{Open} \\ \midrule
AWF~\cite{tong2026awf} & Worst-Case Mutual Information & Learned Primal--Dual Model & Sizes; Constraints; Modulations & Both \\
Emery et al.~\cite{emery2025massivemimoprecoding} & Rate--Power Trade-Off & Feature Extractor & Cross-Site Precoding & / \\
SiFo~\cite{zhao2026sifo} & CSI-Capture Efficiency & SSB Codebook; Subspace Predictor & Target-Site CSI Feedback & / \\
WiFo-E~\cite{wen2026wifoe} & Expected Sum Rate & Shared Trunk & Unseen FDD Configurations & / \\ \bottomrule
\end{tabularx}\vspace{-4mm}
\end{table*}

Direct utility optimization aligns model decisions with communication performance without requiring reference actions, but it can specialize the updated parameters to the assumed channel model, system constraints, and utility definition. Utility losses constitute CFM training only when they update shared parameters that are subsequently reused; when the pretrained backbone is frozen, the same losses instead train a downstream head, decoder, or adapter during adaptation.

\subsubsection{Hybrid and Multi-Objective Pretraining}
\label{sec:hybrid_objectives}

Hybrid pretraining combines learning targets from two or more of the five primary groups when optimizing the shared parameters retained for downstream reuse. Multi-objective training is broader: several losses may span different target groups or refine the same target through auxiliary constraints. A generic joint objective is

\begin{small}
\begin{equation}
\begin{aligned}
\mathcal{L}_{\mathrm{joint}}={}&
\lambda_{\mathrm{obs}}\mathcal{L}_{\mathrm{obs}}
+\lambda_{\mathrm{con}}\mathcal{L}_{\mathrm{con}}
+\lambda_{\mathrm{lat}}\mathcal{L}_{\mathrm{lat}}+\lambda_{\mathrm{tt}}\mathcal{L}_{\mathrm{tt}}
+\lambda_{\mathrm{util}}\mathcal{L}_{\mathrm{util}},
\end{aligned}
\end{equation}\end{small}where the nonnegative weights select and balance the objectives used at a given training stage.

The most common pattern combines observation-space reconstruction with a relational constraint. ContraWiMAE~\cite{guler2025multi} and WiFo-MiSAC~\cite{liu2026wifomisac} add contrastive alignment to masked modeling, while BRCL~\cite{ryu2026brcl} combines CSI reconstruction with beam-response contrastive learning based on transmit-side Gram-matrix similarities. T. Jiao et al.~\cite{jiao2025addressingcursescenariotask} align reconstructed channel features with environmental observations, whereas Yazdnian et al.~\cite{yazdnian2026multimodalwireless} augment multimodal masked reconstruction with physics-guided spatial-spectrum distillation. LWLM~\cite{pan2025lwlm} constructs positioning relations around masked CSI modeling, and AM-FM~\cite{zhu2026amfm} supplements reconstruction with alignment and autocorrelation objectives. These methods retain observation detail while encouraging invariance or consistency, but the useful information preserved by the joint representation depends on how the paired views, physical targets, or spatial relations are constructed.

Other combinations connect reconstruction to latent, teacher-generated, or downstream targets. Buffelli et al.~\cite{buffelli2025foundationcommunication} combine masked modeling of channel statistics with transmission-rank and precoder targets in a shared heterogeneous-token Transformer. GFM-RA~\cite{sheng2026gfmra} combines graph reconstruction with teacher alignment, JEPA-CFM~\cite{gao2026jepacfm} couples channel reconstruction with latent-space prediction, and M3F-UAV~\cite{gao2026m3fuav} combines masked feature modeling with positioning targets. ICWLM~\cite{wen2026icwlm} combines algorithm-generated precoding targets with future-CSI prediction in a shared Transformer. The hierarchical WFM of Y. Wang et al.~\cite{wang2026hierarchicalwfm} pretrains its foundation channel encoder (FCE) through masked CSI reconstruction and trains its foundation optimization decoder (FOD) through supervised warm-up followed by system-utility optimization; the resulting backbones are reused for channel estimation and beam prediction. MUSE-FM~\cite{zheng2025musefmmultitaskenvironmentawarefoundation} instead combines task-target losses with a differentiable sum-rate utility. The single-task BERT4beam formulation~\cite{li2025bert4beam} likewise combines optimization-generated beamformer targets with a differentiable system-utility term, whereas its UBERT variant uses supervised multitask beamformer targets. These objectives inject complementary structural or task information into the retained parameters, although gains from the complete model do not by themselves identify which objective or interaction produces the improvement.

SPA-MAE~\cite{chen2026spamae} provides a recent physics-guided composition with a generic MAE backbone. In addition to masked CSI reconstruction, the retained encoder predicts a sparse transformed-domain structure target and aligns its summary token with features extracted by a parameter-aware branch from simulated multipath delay, power, angle, and path-loss values. Ablations of the two auxiliary losses indicate improvements across downstream tasks, including frozen-encoder evaluation. Because the physical-prior branch is removed after pretraining, the current results provide evidence that physical supervision is useful without assigning a direct physical meaning to individual latent coordinates.

Table~\ref{tab:hybrid_methods} summarizes methods that jointly apply objectives from different groups to the retained representation.

\begin{table*}[t]
\caption{Representative hybrid and multi-objective CFM pretraining methods.}
\label{tab:hybrid_methods}
\centering
\fontsize{6.2}{7.2}\selectfont
\setlength{\tabcolsep}{0.8pt}
\renewcommand{\arraystretch}{1.03}
\begin{tabularx}{\textwidth}{@{}L{1.95cm}llY C{0.6cm}@{}}
\toprule
\textbf{Model} & \textbf{Combined targets} & \textbf{Reused component} & \textbf{Downstream tasks} & \textbf{Open} \\ \midrule
AM-FM~\cite{zhu2026amfm} & Reconstruction; alignment; autocorrelation & Wi-Fi CSI encoder & Nine Wi-Fi Sensing Tasks & / \\
BRCL~\cite{ryu2026brcl} & CSI reconstruction; beam-response contrast & CSI encoder & Beam Prediction; User Selection; Future Beam Prediction & / \\
Buffelli et al.~\cite{buffelli2025foundationcommunication} & Channel-statistic reconstruction; rank/precoder targets & Transformer & Rank/Precoder Selection; Doppler-Spread/Delay-Profile Estimation & / \\
ContraWiMAE~\cite{guler2025multi} & Reconstruction; contrastive alignment & MAE encoder & Beam Prediction; LoS/NLoS Classification; Channel Estimation & Code \\
GFM-RA~\cite{sheng2026gfmra} & Graph reconstruction; teacher alignment & Graph encoder & Power Control & / \\
ICWLM~\cite{wen2026icwlm} & Precoding targets; future-CSI prediction & Shared Transformer & In-Context Precoding; Channel Extrapolation & / \\
JEPA-CFM~\cite{gao2026jepacfm} & Reconstruction; latent prediction & Fluid-antenna encoder & Channel Extrapolation; Positioning & / \\
LWLM~\cite{pan2025lwlm} & CSI reconstruction; positioning relations & Positioning encoder & Time of Arrival (ToA) and AoA Estimation; Positioning & Code \\
M3F-UAV~\cite{gao2026m3fuav} & Feature modeling; position targets & Multimodal backbone & Positioning; Beam Prediction; Time-Domain Channel Extrapolation & / \\
Jiao et al.~\cite{jiao2025addressingcursescenariotask} & Channel reconstruction; cross-modal alignment & Dual encoders + fusion & CSI Feedback; Positioning; Beam Prediction; LoS/NLoS Classification & / \\
Yazdnian et al.~\cite{yazdnian2026multimodalwireless} & Reconstruction; spatial-spectrum distillation & Multimodal encoders & Positioning; Beam Prediction & / \\
MUSE-FM~\cite{zheng2025musefmmultitaskenvironmentawarefoundation} & Task targets; sum-rate utility & Shared Transformer & Channel Estimation; MIMO Detection; Channel Decoding; Beamforming; Positioning & / \\
SPA-MAE~\cite{chen2026spamae} & Reconstruction; sparse structure; multipath-feature alignment & MAE encoder & LoS/NLoS Classification; Channel Estimation; Beam Prediction; Positioning & / \\
Y. Wang et al.~\cite{wang2026hierarchicalwfm} & Reconstruction; task targets; system utility & FCE/FOD backbones & Channel Estimation; Beam Prediction & / \\
WiFo-MiSAC~\cite{liu2026wifomisac} & Multimodal reconstruction; contrastive alignment & MoE encoder & Frequency-Domain Channel Extrapolation; Channel Estimation; Beam Prediction; Distance Estimation; AoA Estimation & / \\ \bottomrule
\end{tabularx}\vspace{-4mm}
\end{table*}

Hybrid and multi-objective pretraining concerns learning objectives applied to parameters that are subsequently retained for reuse. These objectives may be optimized jointly or in successive phases of base-model construction, whereas architectural hybridization and multimodal fusion describe model design. Existing CFM studies most frequently combine reconstruction with relational learning, while combinations involving task targets, latent prediction, and system utilities are less common. Matched single-objective ablations, loss-weight sensitivity, additional training cost, and analysis of complementarity, gradient conflict, or task-specific degradation can reveal whether the combined objectives provide consistent benefits~\cite{yu2020gradientsurgery}.

\subsection{Adaptation and Transfer in Existing CFM Studies}
\label{sec:existing_adaptation}

The pretraining objective does not uniquely determine the downstream adaptation strategy. Full fine-tuning and frozen-backbone head or decoder training are the most common choices. CSI-MAE~\cite{jiang2026csimaemaskedautoencoderbasedchannel}, for example, evaluates both full fine-tuning and a frozen encoder with an adapted decoder for channel extrapolation. CSI2Vec~\cite{palhares2025csi2vec} uses scenario-specific heads, WiFo-E~\cite{wen2026wifoe} adapts configuration-specific branches attached to a shared trunk, and RA-LWLM~\cite{pan2026ralwlm} combines a frozen encoder with a target-scene fingerprint database.

The Routing Adapter for Feature Composition (RAFC)~\cite{shi2026rafc} provides a distinct parameter-efficient interface. It freezes the pretrained backbone, summarizes selected intermediate hidden states into compact descriptors, and learns task- and sample-dependent routing weights that combine the corresponding full-resolution features. Across the reported backbones, the trainable router contains 0.003--0.049 million parameters. RAFC therefore adapts which representation depths a task uses, rather than updating the backbone or relying only on its final layer.

PEFT and direct inference are less frequently examined. OmniLoc~\cite{chu2026omniloc} compares zero-shot inference, frozen-backbone head tuning, and LoRA for cross-environment positioning. Direct inference additionally requires the pretrained output interface to match the new task: direct multi-step forecasting has been applied to an unseen ISAC time-delay variable~\cite{sheng2025wirelessfoundationmodelmultitask}, whereas phys-WFM~\cite{wang2026physwfm} examines zero-shot behavior under controlled physical transformations.

Transfer may change the task, propagation scenario, deployment site, or radio configuration. CSI2Vec illustrates cross-task reuse, SiFo~\cite{zhao2026sifo} evaluates unseen-site CSI feedback, and WiFo-E studies heterogeneous FDD configurations. Parameter updates and target-domain information are separate dimensions: a decoder trained on labeled target samples performs adaptation even with a frozen backbone, while a fixed encoder can still use a target-domain resource such as a fingerprint database. Accordingly, full fine-tuning, probing, frozen nonlinear heads, PEFT, and direct inference characterize different forms of reuse and are not directly interchangeable.

\subsection{Inference-Efficient CFM Methods}
\label{sec:inference_efficiency}

Inference efficiency constitutes a research stage distinct from pretraining and downstream adaptation. It can be pursued by reducing the portion of a pretrained model executed for each task, compressing the reusable model into a smaller student, or designing a more efficient backbone before pretraining. These approaches change different parts of the CFM pipeline and need not preserve the same parameters or adaptation interface.

Early-exit methods attach lightweight task heads to intermediate layers~\cite{teerapittayanon2016branchynet}. O. Mashaal et al.~\cite{mashaal2026earlyexits} select a task-specific fixed exit after profiling a frozen wireless encoder, while dynamic sample-wise routing is evaluated as a comparator. This setup characterizes layer-depth reuse and task-dependent accuracy--computation trade-offs. Teacher--student compression provides a different route~\cite{hinton2015distilling}: Tiny-WiFo~\cite{zhang2026tinywifo} distills a channel-extrapolation model into a smaller deployed student. Early exit retains the original backbone and shortens its executed path, whereas distillation replaces it with a separately trained model whose transfer properties must be evaluated independently.

Native efficient backbones, including state-space, spiking, and hybrid attention--state-space designs discussed in Section~\ref{sec:backbones}, alter the operator used throughout pretraining and inference rather than compressing an already trained CFM. The approaches therefore preserve different reuse interfaces: early exit retains the original backbone but shortens its executed path, distillation produces a separate student, and an efficient backbone changes the base model itself. Their reported savings describe different components and should be interpreted together with the associated task performance and adaptation setting.

\subsection{Cross-Method Comparison and Key Findings}
\label{sec:method_synthesis}

The five learning-target groups span a shift from learning structure present in channel or signal observations to optimizing predefined task outputs and system objectives. This shift can strengthen alignment with a known use case, but it also introduces more assumptions about tasks, reference outputs, or system models. Table~\ref{tab:objective_synthesis} compares their targets, primary focus, and key design choices. Hybrid methods combine targets represented by two or more rows rather than forming an additional primary group.

\begin{table*}[t]
\caption{Comparison of the five primary CFM learning-target groups.}
\label{tab:objective_synthesis}
\centering
\scriptsize
\setlength{\tabcolsep}{1.5pt}
\renewcommand{\arraystretch}{1.10}
\begin{tabularx}{\textwidth}{@{}L{3.9cm}L{4.2cm}L{4.0cm}Y@{}}
\toprule
\textbf{Learning-target group} & \textbf{Learning target} & \textbf{Primary focus} & \textbf{Key design choice} \\ \midrule
Observation-Space Modeling & Masked, noisy, or future observations & Observation fidelity and prediction & Context, corruption, and target design \\ \addlinespace[0.8pt]
Relational Representation Learning & Relations across associated views & Alignment and selected invariance & Relation semantics and collapse control \\ \addlinespace[0.8pt]
Latent-Space Prediction & Target-encoder representations & Predictable latent structure & Context--target scope and target-encoder update \\ \addlinespace[0.8pt]
Task-Target Learning & Labels and reference outputs & Predefined-output alignment & Target quality, task mix, and loss balance \\ \addlinespace[0.8pt]
System-Utility Optimization & Utility or feedback reward & End-to-end decision quality & System model, utility, and constraints \\ \bottomrule
\end{tabularx}\vspace{-4mm}
\end{table*}

The learning target determines which information the reusable model is explicitly trained to preserve. Observation-space objectives retain coefficient- or signal-level detail and are therefore closely matched to reconstruction, interpolation, and extrapolation~\cite{liu2025wifo,jiang2026csimaemaskedautoencoderbasedchannel}; they may also retain hardware-, waveform-, or preprocessing-specific variation. Relational objectives use cross-view, temporal, spatial, or cross-modal associations to select invariances~\cite{jiang2025mimo,salihu2024selfsupervised}, but incorrect relations or aggressive augmentations can suppress factors needed by a later task. Latent prediction depends on its context--target construction~\cite{assran2023selfsupervised,gao2026jepacfm}, while task-target and utility objectives align more directly with specified outputs or system performance and consequently depend more strongly on the task mix, reference generator, channel model, constraints, and utility.

Taken together, Table~\ref{tab:objective_synthesis} suggests that objective selection is tied to the intended reuse interface rather than a universal hierarchy. Observation-space modeling provides a direct match when downstream modules require coefficient-level reconstruction or prediction. Relational and latent-space objectives can suit a shared encoder supporting heterogeneous heads, provided that the selected relations or context--target pairs preserve the channel factors required by those heads. Task-target learning fits a stable family of known outputs, whereas utility optimization directly addresses a specified system decision. Moving toward task or utility alignment can reduce the gap between pretraining and a known application, but does not imply broader transfer to unseen tasks or radio conditions. Sharing the same backbone or pretraining data also does not make objectives equivalent because their targets can reward different information. Supervision type alone likewise does not determine reuse breadth: self-supervised targets may encode narrow assumptions, while supervised targets may support multiple tasks or operating conditions. The relevant downstream test is whether the retained representation supports the intended tasks when the task or operating conditions change.

Current results do not establish a universally preferable objective family. The choice depends on the required output, the factors to retain or suppress, the available supervision, and the expected deployment shift. Hybrid training can combine observation fidelity, relational structure, and task or utility alignment, but its benefit requires matched objective and interaction ablations because additional computation, loss imbalance, gradient conflict, and degradation on individual tasks may otherwise be hidden~\cite{yu2020gradientsurgery}.

The adaptation protocol changes what a downstream result demonstrates. Frozen-backbone heads test reuse of a fixed representation, PEFT tests restricted parameter updates, full fine-tuning tests the pretrained initialization under a larger update budget, and direct inference requires a compatible pretrained interface. These protocols characterize different forms of generalization. Meaningful comparisons therefore depend on the target-data budget, trainable parameters, access to target-domain resources, and the task or operating conditions being changed.

Tables~IV--IX also show why current results support conditional gains rather than a cross-paper ranking. Pretraining data, backbone capacities, downstream heads, simulation or measurement settings, target-domain exposure, and metrics differ widely across studies. A defensible performance comparison therefore controls these factors and includes matched from-scratch and objective-ablation baselines~\cite{10288574}. Claims of channel-aware inductive bias additionally require matched architectural ablations and unseen physical conditions; simulation-only gains do not establish robustness on measured channels.

\section{Applications of CFMs}
The CFM scope extends beyond the physical layer (PHY). A channel representation may directly reconstruct or predict CSI, or it may support precoding, scheduling, beam management, positioning, sensing, or radio-map estimation. The relevant distinction is whether pretrained parameters or outputs are reused for a channel-related task, rather than which protocol layer contains that task. Whereas Section~\ref{sec:landscape} organizes studies by their methods, this section compares their applications, evaluation metrics, and system constraints.

\subsection{CFM Applications and Evaluation Metrics}

Table~\ref{tab:application_interfaces} compares representative outputs, metrics, and evaluation settings across four application groups.

Representative studies illustrate how the pretraining methods in Section~III support the applications summarized in the table. PilotWiMAE~\cite{guler2026pilotwimae} pretrains an encoder on noisy pilots through masked observation-space modeling and subsequently attaches channel-estimation or cross-frequency beam-prediction heads. GFM-RA~\cite{sheng2026gfmra} represents channel gains as an interference graph, pretrains a graph encoder through edge reconstruction and teacher alignment, and adapts the encoder to resource-allocation objectives. WWM~\cite{chen2026wirelessworldmodel} performs latent prediction from CSI, point clouds, and trajectories before attaching heads for several channel-related tasks.

\begin{table*}[t]
\caption{Representative CFM applications and evaluation criteria by primary system role.}
\label{tab:application_interfaces}
\centering
\scriptsize
\setlength{\tabcolsep}{1.2pt}
\renewcommand{\arraystretch}{1.08}
\begin{tabularx}{\textwidth}{@{}L{3.9cm}L{4.15cm}L{3.6cm}Y@{}}
\toprule
\textbf{Application} & \textbf{Representative output} & \textbf{Primary metrics} & \textbf{Main evaluation settings} \\ \midrule
\multicolumn{4}{@{}l}{\textbf{Physical-Layer Processing}} \\
Channel Estimation and Interpolation & Estimated or interpolated CSI & NMSE & SNR; pilot pattern or density \\
Channel Extrapolation & Extrapolated CSI & NMSE & SNR; extrapolation domain; prediction ratio or horizon \\
CSI Feedback & Compressed or reconstructed CSI & NMSE & Compression ratio; quantization bits \\
Signal Detection and Decoding & Detected symbols or decoded bits & SER/BER & SNR; modulation and coding; user count \\
Precoding & Precoding matrix or beamformer & Sum rate & User and antenna counts \\
Channel-Attribute Inference & Channel class or physical parameter & Accuracy/F1; MAE & SNR \\
\addlinespace[0.5pt]\cmidrule(lr){1-4}
\multicolumn{4}{@{}l}{\textbf{Channel-Dependent RAN and MAC Tasks}} \\
Beam Prediction & Beam index or ranking & Top-$k$ accuracy & SNR; codebook size \\
Resource Allocation & Schedule, power, or assignment & Normalized utility ratio & Network size; utility objective \\
\addlinespace[0.5pt]\cmidrule(lr){1-4}
\multicolumn{4}{@{}l}{\textbf{ISAC}} \\
Positioning & Estimated 2D or 3D position & Mean/percentile error & Frequency band \\
RF Sensing & Activity or environmental attribute & Accuracy & Subject; sensing device \\
\addlinespace[0.5pt]\cmidrule(lr){1-4}
\multicolumn{4}{@{}l}{\textbf{Radio-Environment Modeling}} \\
Radio-Map Modeling & Spatial, temporal, or spectral map & RMSE/NMSE & Sampling rate; conditioning inputs \\
Channel Charting & Low-dimensional channel chart & Trustworthiness/continuity & Embedding dimension; neighborhood size \\ \bottomrule
\end{tabularx}
\vspace{1pt}
\parbox{\textwidth}{\scriptsize\textit{Abbreviations:} NMSE, normalized mean-square error; SER, symbol error rate; BER, bit error rate; MAE, mean absolute error; RMSE, root-mean-square error; SNR, signal-to-noise ratio. Channel-attribute inference covers channel or scenario classes and physical parameters such as path loss, AoA, or Doppler shift.}\textbf{}
\end{table*}

\subsection{Representative Matched Case Study}
\label{sec:application_case_study}

To complement the cross-study survey, the following application subsections compare two representative pretraining objectives under matched conditions: the contrastive objective of CSI-CLIP~\cite{jiang2025mimo} and the masked-reconstruction objective of CSI-MAE~\cite{jiang2026csimaemaskedautoencoderbasedchannel}. Both implementations use a common ViT-Base backbone and a DeepMIMO-derived pretraining dataset containing more than 0.7 million CSI samples from 35 indoor and outdoor scenarios~\cite{alkhateeb2019deepmimo}; Vanilla ViT uses the same backbone but is trained from scratch. Table~\ref{tab:case_target_data} lists the seven target scenarios and task-specific splits. These scenarios are excluded from pretraining but used for supervised downstream adaptation and validation. The results therefore provide a matched illustration of reuse across tasks and scenarios, rather than a general benchmark ranking or a zero-shot comparison.

\begin{table}[!t]
\caption{Downstream target scenarios and task-specific data splits.}
\label{tab:case_target_data}
\centering
\footnotesize
\setlength{\tabcolsep}{2.0pt}
\renewcommand{\arraystretch}{1.06}
\begin{tabular*}{\columnwidth}{@{\extracolsep{\fill}}lcccc@{}}
\toprule
\textbf{Scenario} & \textbf{Environment} & \textbf{Carrier (GHz)} & \textbf{Train} & \textbf{Validation} \\ \midrule
Chicago   & Urban   & 3.5 & 228    & 57 \\
San Diego & Urban   & 3.5 & 1,753  & 439 \\
Charlotte & Urban   & 3.5 & 2,881  & 721 \\
I1\_2p4   & Indoor  & 2.4 & 15,693 & 3,924 \\
I2\_28B   & Indoor  & 28  & 15,702 & 3,926 \\
O1\_3p5B  & Outdoor & 3.5 & 38,712 & 9,679 \\
O1\_60    & Outdoor & 60  & 67,017 & 16,755 \\ \bottomrule
\end{tabular*}\vspace{-4mm}
\end{table}

\subsection{Physical-Layer Processing}

At the PHY, CFMs have been evaluated for channel estimation, interpolation, feedback, time- and frequency-domain extrapolation, signal detection, decoding, and precoding. Masked-reconstruction models such as WiFo~\cite{liu2025wifo}, the scalable masked channel model~\cite{guo2026mcm}, and ChannelGPT~\cite{yu2025channelgpt} reuse or extend pretrained channel representations for channel extrapolation or completion. Other channel-extrapolation studies examine multi-domain generalization and cross-band reconstruction~\cite{gao2026ai,gao2025enabling,zhang2026crossbandcsi}. As discussed in Section~III-C, channel-structure-aware variants may additionally modify the signal domain, token-mixing operator, or positional encoding.

\begin{figure}[!t]
    \centering
    \includegraphics[width=0.9\columnwidth]{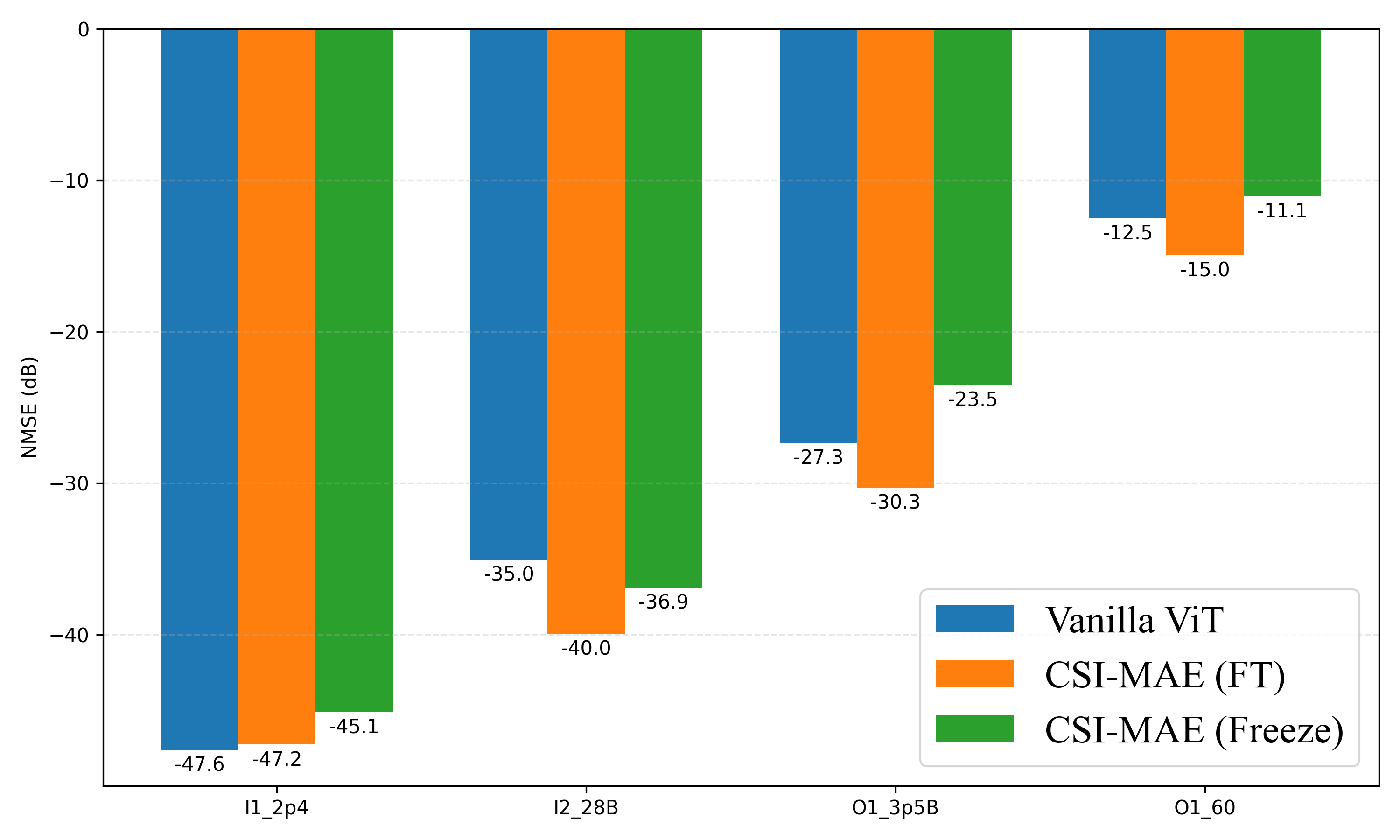}
    \caption{Subcarrier-extrapolation NMSE in four matched target scenarios. Lower (more negative) NMSE is better. FT denotes full fine-tuning, whereas Freeze retains the pretrained CSI-MAE backbone and adapts only the task decoder.}
    \label{fig:case_extrapolation}\vspace{-4mm}
\end{figure}

Fig.~\ref{fig:case_extrapolation} compares subcarrier-extrapolation NMSE for Vanilla ViT and CSI-MAE under full fine-tuning and frozen-backbone adaptation. After full fine-tuning, CSI-MAE achieves lower NMSE than Vanilla ViT in three of four scenarios. Frozen-backbone adaptation updates only the task decoder, approximately 3\% of the parameters, but achieves lower NMSE than Vanilla ViT in only one scenario. The comparison illustrates a trainable-parameter/performance trade-off rather than a uniform advantage of parameter freezing.

Receiver processing provides a different application. MUSE-FM~\cite{zheng2025musefmmultitaskenvironmentawarefoundation} jointly trains modules for channel estimation, MIMO detection, decoding, positioning, and precoding by combining task targets with a differentiable sum-rate objective. FM-Receiver~\cite{zheng2026fmreceiver} instead maps received resource-grid observations to recovered bits through a configuration-adaptive unified inner/outer receiver. These studies extend CFM use beyond an encoder with a single prediction head. However, their end-to-end performance reflects the combined effects of the pretrained encoder, task-conditioned modules, supervised targets, and receiver design. Meaningful cross-configuration comparisons therefore use matched receiver baselines under the same transmission settings.

For precoding, WiFo-E~\cite{wen2026wifoe} reuses a shared trunk, adapts configuration-specific branches, and optimizes a differentiable expected sum-rate objective. MUSE-FM~\cite{zheng2025musefmmultitaskenvironmentawarefoundation} and the model of Y. Wang et al.~\cite{wang2026hierarchicalwfm} likewise connect a channel encoder to task-conditioned optimization modules. Direct utility optimization can align model outputs with system performance without imitating solutions from an iterative solver, but it may also specialize the updated parameters to the assumed channel model and system objective. Evaluation therefore distinguishes encoder reuse from optimizer reuse and reports feasibility, rate--fairness trade-offs, and adaptation to new array and user configurations.

\FloatBarrier
\subsection{Channel-Dependent RAN and MAC Tasks}

Codebook-based beam prediction is the most frequently reported channel-dependent RAN task among the reviewed studies. Operational beam management additionally encompasses initial access, tracking, and recovery under mobility and blockage~\cite{xue2024beammanagement}. LWM~\cite{alikhani2024large}, CSI-CLIP~\cite{jiang2025mimo}, WiMamba~\cite{raviv2026wimamba}, ConsisFormer~\cite{wang2026consisformer}, and WWM~\cite{chen2026wirelessworldmodel} evaluate beam prediction together with other tasks. Their reusable representations may encode spatial, blockage, or cross-band relations, which a classification or ranking head maps to a beam codebook. Table~\ref{tab:case_beam_prediction} shows numerically higher reported Top-1 accuracies for CSI-MAE and CSI-CLIP than for Vanilla ViT in each matched target scenario, with differences ranging from less than one percentage point to more than 19 percentage points. The best result in each scenario is shown in bold.

\FloatBarrier
\begin{table}[t]
\caption{Top-1 beam-prediction accuracy after full fine-tuning on the matched validation splits (\%).}
\label{tab:case_beam_prediction}
\centering
\footnotesize
\setlength{\tabcolsep}{3.0pt}
\renewcommand{\arraystretch}{1.08}
\begin{tabular*}{\columnwidth}{@{\extracolsep{\fill}}lccc@{}}
\toprule
\textbf{Scenario} & \textbf{Vanilla ViT} & \textbf{CSI-MAE} & \textbf{CSI-CLIP} \\ \midrule
Chicago   & 82.46 & 87.72 & \textbf{89.47} \\
San Diego & 68.77 & 73.84 & \textbf{75.85} \\
Charlotte & 78.03 & \textbf{83.60} & 80.86 \\
I1\_2p4   & 95.02 & 95.08 & \textbf{95.36} \\
I2\_28B   & 79.75 & \textbf{83.77} & 83.32 \\
O1\_3p5B  & 92.91 & 93.10 & \textbf{93.18} \\
O1\_60    & 75.29 & 90.77 & \textbf{94.60} \\ \bottomrule
\end{tabular*}\vspace{-4mm}
\end{table}

A complete evaluation nevertheless supplements top-$k$ accuracy with beam-sweeping overhead, robustness to codebook and array changes, outage probability, and latency. A gain in classification accuracy alone does not establish an operational beam-management benefit.

Channel representations can also support scheduling, power control, and interference-aware optimization. The model of Y. Wang et al.~\cite{wang2026hierarchicalwfm} couples its channel encoder to beamforming and user-scheduling modules, whereas GFM-RA~\cite{sheng2026gfmra} reconstructs masked channel-gain edges and aligns graph views before adaptation to wireless resource-allocation objectives. These methods fall within the CFM scope when the pretrained encoder represents channel gains or interference relations and is reused under new propagation or network conditions. They do not imply coverage of traffic, queueing, policy constraints, or the complete RAN state. Cross-topology and cross-objective evaluations remain necessary.

For mobility management, the channel-extrapolation, beam-tracking, blockage, or positioning outputs of models such as LWM-Temporal~\cite{alikhani2026lwmtemporal} and WWM~\cite{chen2026wirelessworldmodel} can serve as inputs to a handover controller. Existing evaluations, however, remain focused on these component tasks rather than the complete mobility-control loop.

\FloatBarrier
\subsection{ISAC}

\begin{figure}[!t]
    \centering
    \includegraphics[width=\columnwidth]{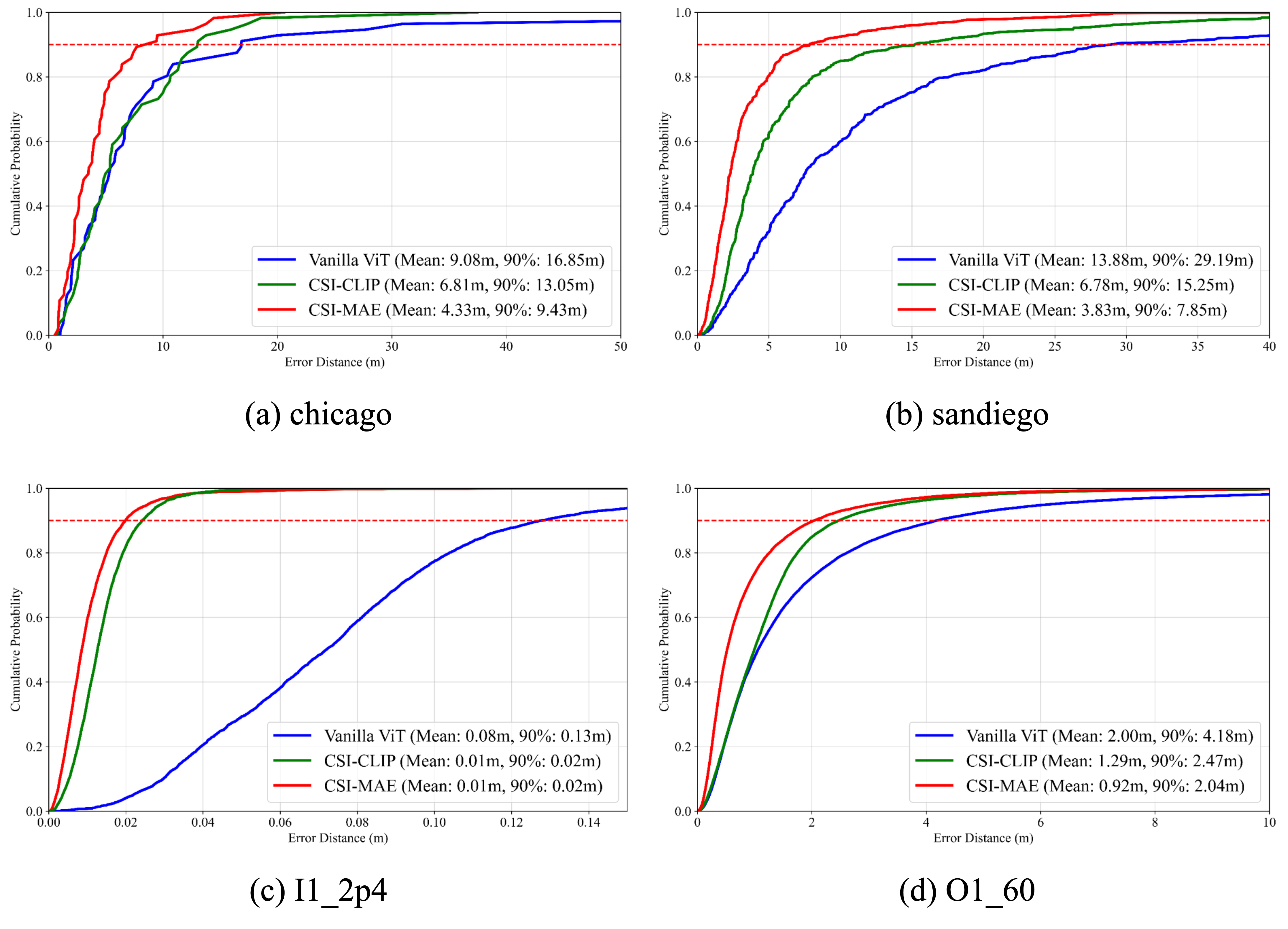}
    \caption{Empirical cumulative distribution functions (CDFs) of positioning error in four matched target scenarios. The horizontal dashed line marks a cumulative probability of 0.9, and the legends report the mean and 90th-percentile errors. The horizontal-axis ranges differ across panels and may truncate extreme error tails.}
    \label{fig:case_positioning}\vspace{-4mm}
\end{figure}

ISAC applications exploit the common dependence of communication, positioning, and sensing on multipath propagation, while retaining task-specific outputs and metrics~\cite{liu2022isac,gao2024performance,gao2025stochastic}. Angle, delay, and Doppler estimation are treated as ISAC tasks when they are inferred from sensing echoes or jointly designed communication--sensing observations; the same quantities inferred from communication-only CSI are instead channel attributes. Positioning is the most frequently studied sensing-related application among existing CFMs. CSI-CLIP~\cite{jiang2025mimo}, CSI-MAE~\cite{jiang2026csimaemaskedautoencoderbasedchannel}, CSI2Vec~\cite{palhares2025csi2vec}, WWM~\cite{chen2026wirelessworldmodel}, WMFM~\cite{farzanullah2025wmfm}, LatentWave~\cite{mohamed2026latentwave}, Map as a Prompt~\cite{chu2026sigmap}, and OmniLoc~\cite{chu2026omniloc} adapt channel representations or multimodal spatial-signal representations to position estimation. Fig.~\ref{fig:case_positioning} shows lower mean and 90th-percentile errors for CSI-CLIP and CSI-MAE than for the task-specific ViT baseline across all four matched target scenarios. CSI-MAE obtains the lowest or tied-lowest mean error in these scenarios, while the magnitude of the gain varies with the target environment and data regime. RA-LWLM~\cite{pan2026ralwlm} instead combines a frozen channel encoder with a target-scene fingerprint database and an in-context positioning module.

\FloatBarrier

Beyond positioning, existing studies connect CFM pretraining to RF sensing and sensing-assisted communication. 6G WavesFM~\cite{aboulfotouh20256g} reuses a shared encoder across human-activity sensing, RF classification, positioning, and channel estimation. WiFo-MiSAC~\cite{liu2026wifomisac} jointly processes CSI, radar, and map data for channel extrapolation and estimation, beam prediction, distance estimation, and angle-of-arrival (AoA) estimation. The multimodal model of Yazdnian et al.~\cite{yazdnian2026multimodalwireless} combines CSI with scene information and evaluates positioning and hybrid precoding. Other multimodal CFMs primarily use sensing context to improve communication tasks: Jiao et al.~\cite{jiao2025addressingcursescenariotask} study feedback, positioning, beam prediction, and LoS/NLoS identification, whereas WiFo-M$^2$~\cite{zhang2026wifom2} uses temporal cross-modal contrast to transfer environmental sensing features to beam prediction, channel estimation and interpolation, channel extrapolation, and positioning.

These studies reflect two complementary uses of CFMs in ISAC. The first reuses a shared channel representation for both communication and sensing outputs, whereas the second introduces environmental sensing as auxiliary information for channel-related communication tasks.

At the downstream interface, an observation-reconstruction encoder can connect to a coefficient-level decoder for channel estimation or extrapolation and to regression or classification heads for position, angle, delay, Doppler, or activity. Joint communication--sensing adaptation may share the encoder while retaining task-specific heads and losses~\cite{aboulfotouh20256g,liu2026wifomisac}. Its benefit should be isolated through single-task heads, separate-encoder baselines, frozen-backbone and full-fine-tuning variants, and metrics for both functions; otherwise, gains may arise from task supervision or auxiliary modalities rather than the pretrained channel representation.

\FloatBarrier
\subsection{Radio-Environment Modeling}

Radio-map modeling is treated separately from PHY processing, RAN/MAC operation, and ISAC because its output is a region-level representation of the radio environment rather than a link-level estimate, resource decision, or sensing result. FM-RME~\cite{yang2026fmrme} estimates spatial, temporal, and spectral radio maps through masked multidimensional modeling, whereas FARM~\cite{gao2026farm} combines sparse radio observations and environmental priors for condition-free and condition-based aerial radio-map construction. Such maps can support network planning, positioning, resource allocation, or sensing-assisted communication.

Channel charting provides a complementary representation of the radio environment by embedding high-dimensional CSI into a low-dimensional chart that preserves relative channel neighborhoods rather than estimating absolute positions or a spatial field. CSI2Vec~\cite{palhares2025csi2vec} and WALoMA~\cite{makin2026waloma} evaluate channel charting as a downstream use of pretrained channel representations.

\FloatBarrier
\section{Challenges and Future Research}
Experimental settings vary widely across CFM studies, limiting general conclusions about transfer, cost, and reliability. The following directions target these evidence gaps.

\subsection{Transfer Evaluation Across Tasks and Operating Conditions}
Three complementary settings are relevant to transfer evaluation: cross-task reuse, cross-condition reuse, and their combination. Cross-task tests change the downstream output while keeping the operating conditions fixed, whereas cross-condition tests retain the task but change the propagation environment or radio configuration. Fair comparisons hold the target-data budget, trainable-parameter budget, downstream head, and matched from-scratch baseline constant, allowing differences to be attributed to pretraining rather than adaptation capacity.

Cross-condition evaluation can examine both individual and combined changes in operating conditions. Direct inference, use of target-domain resources such as a fingerprint database, unsupervised calibration, and supervised adaptation expose the model to different target information and therefore constitute separate protocols. Performance across target-data and trainable-parameter budgets then indicates when pretrained reuse remains useful.

\subsection{Model Advancements}
Future work can evaluate four complementary dimensions:
\begin{itemize}
    \item \textbf{End-to-end efficiency:} Sparse attention, token reduction, state-space models, quantization, and conditional computation should be compared through accuracy, latency, peak memory, energy, preprocessing, task-head cost, and hardware conditions.
    \item \textbf{Model and data scaling:} General-purpose scaling laws need not transfer to CFMs; controlled studies should vary capacity, data diversity, computation, and downstream tasks together~\cite{hoffmann2022computeoptimal}.
    \item \textbf{Propagation-aware design:} Channel structure, geometry, array configuration, and physical constraints can enter through representation, positional encoding, architecture, or objective. Delay--angle transformations, structured masks, or equivariant objectives encode selected priors; they do not by themselves guarantee compliance with Maxwell's equations, identify a new propagation mechanism, or produce a closed-form channel model. Each claim should identify the encoded property and test it through matched architectural ablations, controlled physical transformations, feasibility residuals where available, and unseen, preferably measured, data~\cite{xiao2025wireless}.
    \item \textbf{Physical representation tests, interpretability, and reliability:} Downstream accuracy does not establish that an embedding exposes propagation variables. Frozen probes can test whether path loss, delay spread, angular spread, K-factor, AoA, ToA, or Doppler are accessible, while controlled interventions can test sensitivity or invariance when one factor changes. Probe data should be separated from pretraining and adaptation data, with untrained and task-specific encoders as references. Existing results such as path-loss transfer in SWiT~\cite{salihu2024selfsupervised} and multipath-parameter supervision in SPA-MAE~\cite{chen2026spamae} provide useful evidence, but neither downstream prediction nor auxiliary physical supervision assigns a physical meaning to individual latent coordinates. Uncertainty should be tested under the same distribution shifts~\cite{ovadia2019uncertainty}.
\end{itemize}

\subsection{Data Coverage and Reproducible Evaluation}
Data requirements depend on modality, model capacity, downstream use, and adaptation setting. Reports should document environment and configuration coverage, preprocessing, and separation of pretraining, adaptation, validation, and testing. Simulated data provide scale and control; measured data capture hardware and site effects. Mixing them does not by itself demonstrate zero-shot simulation-to-measurement transfer.

Beyond the data resources in Table~\ref{tab:pretraining_data}, a benchmark specifies tasks and evaluation procedures. Two recent resources provide complementary controls. The LWM Challenge 2025~\cite{lwmchallenge2025} covers five channel tasks and evaluates complete systems through organizer retraining, task-data limits, and hidden tests, while allowing external pretraining data. CFM-Bench~\cite{gao2026cfmbench} covers six domains and six task groups, fixes adaptation protocols and splits, excludes benchmark samples from pretraining, and isolates physically related test units. Neither prescribes a common pretraining dataset. Future reports should additionally state target-domain exposure, adaptation budget, baselines, repeated-run variation, measured-domain results, computation, and whether target measurements calibrate a simulation-to-measurement experiment.

\subsection{Deployment Cost, Model Updating, Privacy, and Security}

Deployment cost is not captured by parameter count or offline accuracy. A frozen CFM still incurs inference cost, while test-time adaptation adds update latency~\cite{cheng2026howbigwfm}. Evaluations should report end-to-end latency, peak memory, energy, precision, hardware, and radio deadlines together with accuracy after compression or adaptation.

Model updates after deployment should be specified as an executable procedure. For changes in carrier frequency, array geometry, calibration, mobility, or interference, studies should report how the change is detected, how many labeled or unlabeled samples are collected, which parameters are updated, and the gradient steps and update time. The model should then be retested on earlier conditions, with rollback tied to increased outage or beam-failure rate. Deployment-time model updating remains underexplored in current CFM research.

Channel measurements can expose locations, trajectories, occupancy, device identities, or site characteristics, while released weights or embeddings can introduce membership-inference risks~\cite{wiesmayr2025caez,zhang2026wiloc,shokri2017membership}. Distributed pretraining keeps raw measurements local but does not prevent gradient leakage, poisoning, or backdoors~\cite{zhu2019deepleakage,bagdasaryan2020backdoor}. Evaluations should define the protected information and attacker access, test data, updates, embeddings, and weights, and report privacy or robustness costs with task performance. Direct CFM evidence remains limited.

\vspace{-4mm}
\section{Conclusions}
This survey defines CFMs through channel-representation pretraining and reuse, independently of modality, architecture, scale, or protocol layer. Its taxonomy covers observation-space, relational, latent-predictive, task-target, system-utility, and hybrid learning, with data, backbone, adaptation, application, and evaluation as complementary dimensions.

Current evidence supports conditional reuse, not universal generalization. Simulation-heavy data, differing adaptation protocols, and limited measured-domain and physical-representation tests call for matched baselines, reproducible benchmarks, and joint assessment of accuracy, efficiency, model updating, reliability, privacy, and robustness. Interpretability additionally requires frozen probes and controlled changes to propagation factors under unseen conditions, compared with untrained and task-specific representations; downstream accuracy alone does not establish physically meaningful internal variables.

\bibliographystyle{IEEEtran}
\bibliography{bibfile_revision}

\end{document}